\documentclass[11pt]{article}

\usepackage{graphicx,amsmath,amsmath,amsfonts,amssymb,dcolumn,amsthm,euscript}

\begin{document}

%




\date{}
\title{\textbf{Aspects of the refined Gribov-Zwanziger action in linear covariant gauges
}}


 \author{\textbf{M.~A.~L.~Capri}\thanks{caprimarcio@gmail.com}$\,\,\,\,^{a}$\,\,,
 \textbf{D.~Fiorentini}\thanks{diegofiorentinia@gmail.com}$\,\,\,\,^{a}$\,\,,
 \textbf{A.~D.~Pereira}\thanks{aduarte@if.uff.br}$\,\,\,\,^{a,b}$\,\,,
  \\
  \textbf{R.~F.~Sobreiro}\thanks{sobreiro@if.uff.br}$\,\,\,\,^{b}$\,\,,
 \textbf{S.~P.~Sorella}\thanks{silvio.sorella@gmail.com}$\,\,\,\,^{a}$\,\,,
 \textbf{R.~C.~Terin}\thanks{rodrigoterin3003@gmail.com}$\,\,\,\,^{a}$
 \\[2mm]
 {\small \textnormal{ \it $^{a}$  UERJ -- Universidade do Estado do Rio de Janeiro,}}\\
 {\small \textnormal{ \it Instituto de F\'{\i}sica, Departamento de F\'{\i}sica Te\'{o}rica,}}\\
 {\small \textnormal{ \it Rua S\~{a}o Francisco Xavier 524, 20550-013 Maracan\~{a}, Rio de Janeiro, RJ, Brasil}}
  \\[2mm]
 {\small   \textnormal{  \it $^{b}$ UFF -- Universidade Federal Fluminense,}}\\
{\small \textnormal{ \it Instituto de F\'{\i}sica, Campus da Praia Vermelha,}}\\
\small \textnormal{ \it Avenida General Milton Tavares de Souza s/n, 24210-346, Niter\'oi, RJ, Brasil}\\
} 	

\maketitle


\begin{abstract}


\noindent We prove the renormalizability to all orders  of a refined Gribov-Zwanziger type action in linear covariant gauges in  four-dimensional Euclidean space. In this model, the Gribov copies are taken into account by requiring  that the Faddeev-Popov operator is positive definite with respect to the transverse component of the gauge field, a procedure which turns out to be analogous to the restriction to the Gribov region in the Landau gauge. The model studied here can be regarded as the first approximation of a more general nonperturbative BRST invariant formulation of the refined Gribov-Zwanziger action in linear covariant gauges obtained recently in \cite{Capri:2015ixa,Capri:2016aqq}. A key ingredient of the set up worked out in  \cite{Capri:2015ixa,Capri:2016aqq} is the introduction of a gauge invariant  field configuration $\mathbf{A}_{\mu}$ which can be expressed as an infinite non-local series in the starting gauge field $A_\mu$.  In the present case,  we consider the approximation in which only the first term of the series representing $\mathbf{A}_{\mu}$ is considered, corresponding to a pure transverse gauge field. The all order renormalizability of the resulting action  gives thus a strong  evidence of the renormalizability of the aforementioned  more general nonperturbative BRST invariant formulation of the Gribov horizon in linear covariant gauges. 
\end{abstract}

\section{Introduction}

The Gribov-Zwanziger framework, originally  developed in the Landau gauge, enables us to take into account the existence of zero-modes of the Faddeev-Popov operator in the quantization of Euclidean Yang-Mills theories.  The existence of such zero-modes was  pointed out by Gribov \cite{Gribov:1977wm}, who showed that, for a given gauge field $A^{a}_\mu$ which satisfies the Landau gauge condition, {\it i.e.} $\partial_{\mu}A^{a}_{\mu}=0$, there are  other gauge field configurations $A'^{a}_{\mu}$ which are  related to $A^{a}_{\mu}$ via an infinitesimal gauge transformation \textit{and} which also satisfy the Landau gauge condition if the Faddeev-Popov operator develops zero-modes. Hence, the very existence of such zero-modes is associated with the fact that the same gauge orbit is crossed more than once by the gauge-fixing section. Such spurious configurations are the so-called Gribov copies and their existence characterizes the Gribov problem\footnote{For a pedagogical introduction to the Gribov problem, see \cite{Sobreiro:2005ec,Vandersickel:2012tz,Vandersickel:2011zc,Pereira:2016inn}.}. Although this is not a particular feature of the Landau gauge-fixing condition \cite{Singer:1978dk}, this is the gauge in which the Gribov phenomenon is better understood. In the Landau gauge, the gauge field is  purely transverse, {\it i.e.} $\partial_{\mu}A^{a}_{\mu}=0$, a property which ensures that the Faddeev-Popov operator,
\begin{equation}
\mathcal{M}^{ab}(A)=-\partial_{\mu}D^{ab}_{\mu}(A)=-\partial_{\mu}(\delta^{ab}\partial_{\mu}-gf^{abc}A^{c}_{\mu})\,,   \label{mfp}
\end{equation}
is Hermitian. In order to take into account the existence of such copies, Gribov proposed to restrict the path integral domain  to a certain region $\Omega$, known as the Gribov region, in which the Faddeev-Popov operator is positive\footnote{Notice here that the Hermiticity property is fundamental in order to define the Gribov region in terms of positive eigenvalues of the Faddeev-Popov operator.}, namely, 
\begin{equation}
\Omega=\big\{\,A^{a}_{\mu}\,\,\big|\,\,\partial_{\mu}A^{a}_{\mu}=0\,,\,\,\mathcal{M}^{ab}(A)>0\,\big\}\,.
\end{equation}
 It is possible to show that this region enjoys many important properties \cite{Dell'Antonio:1991xt}: $(i)$ it is bounded in all directions in field space; $(ii)$ it is convex;  $(iii)$ every gauge orbit crosses at least once the region $\Omega$. The procedure implemented by Gribov was performed in a semi-classical approximation and has been  generalized later on by Zwanziger to all orders, see \cite{Zwanziger:1989mf}. Although Gribov and Zwanziger followed different strategies, their approaches can be proven to be equivalent, as shown in  \cite{Capri:2012wx}. 

According to \cite{Zwanziger:1989mf},  the restriction of the domain of integration in the path integral  to the Gribov region  $\Omega$  in the Landau gauge is achieved by the addition, to the Faddeev-Popov gauge-fixed Yang-Mills action, of the so-called horizon function, $H(A)$, and of a vacuum term, namely
\begin{eqnarray}
S^{H}_{\mathrm{Landau}}&=&\gamma^4\,H(A)-4\gamma^4V(N^2-1)\,,\nonumber\\
H(A)&=&g^{2}\int d^{4}xd^{4}y\,f^{abc}A^{b}_{\mu}(x)\,\left[\left(-\partial\cdot D\right)^{-1}\right]^{ad}\!(x,y)\,f^{dec}A^{e}_{\mu}(y)\,,
\label{intro1}
\end{eqnarray} 
with $V$ being the space-time volume and $\gamma^{2}$ a mass parameter, known as the Gribov parameter, which is not free, but determined by the gap equation \cite{Zwanziger:1989mf}
\begin{equation}
\langle H(A)\rangle=4V(N^2-1)\,.
\label{intro2}
\end{equation}
The horizon function is non-local, however it is possible to cast it in a local form by the introduction of a suitable set of auxiliary fields. We will show the explicit form of this localization later on.

Nevertheless, the Hermiticity property of the Faddeev-Popov operator is, in general, lost outside of the Landau gauge. This is the case, for example, of the class of gauges known as the linear covariant gauges, given by the condition $\partial_\mu A^a_\mu = i \alpha b^a$, where $\alpha$ is a non-negative gauge parameter and $b^a$ the Lagrange multiplier field. 

In the usual Faddeev-Popov quantization framework, for the gauge-fixed $SU($N$)$ Yang-Mills action in four dimensional Euclidean space in the linear covariant gauges, we have 
\begin{equation}
S_{\mathrm{FP}}=\int d^{4}x\,\bigg(\frac{1}{4}\,F^{a}_{\mu\nu}F^{a}_{\mu\nu}
+\frac{\alpha}{2}\,b^{a}b^{a}+ib^{a}\,\partial_{\mu}A^{a}_{\mu}
+\bar{c}^{a}\,\partial_{\mu}D^{ab}_{\mu}(A)c^{b}\bigg)\,,     \label{fplcg} 
\end{equation}
where 
\begin{equation}
D^{ab}_{\mu}(A)=\delta^{ab}\partial_{\mu}-gf^{abc}A^{c}_{\mu}\,,
\end{equation}
is the covariant derivative in the adjoint representation of the gauge group. The first term in expression \eqref{fplcg} is the usual Euclidean Yang-Mills action with 
\begin{equation} F^{a}_{\mu\nu}=\partial_{\mu}A^{a}_{\nu}-\partial_{\nu}A^{a}_{\mu}+gf^{abc}A^{b}_{\mu}A^{c}_{\nu}\,.
\end{equation}
The field $b^{a}$ characterizes  the gauge-fixing condition in an off-shell way. In fact, its classical equation of motion gives
\begin{equation}
\frac{\delta S_{\mathrm{FP}}}{\delta b^{a}}=i\partial_{\mu}A^{a}_{\mu}+\alpha\,b^{a}=0\quad\Rightarrow\quad 
\partial_{\mu}A^{a}_{\mu}=i\alpha\,b^{a}\,,
\end{equation}
reproducing thus the gauge-fixing condition. Also, when the gauge parameter $\alpha$ goes to zero, the Landau gauge is recovered as a particular case. Finally, the anti-commuting  fields $\{\bar{c}^{a},c^{a}\}$ are the  Faddeev-Popov ghosts\footnote{By convention $\bar{c}^{a}$ is the antighost and $c^{a}$ the ghost fields.}.

In the last years, several efforts have been undertaken in order to generalize the Gribov-Zwanziger approach to the case of the linear covariant gauges. The first results in this direction were obtained in \cite{Sobreiro:2005vn}, in which the gauge parameter $\alpha$ was considered as an infinitesimal parameter. Later on, an extension to finite $\alpha$ was proposed in \cite{Capri:2015pja}. The main point in these two works is that the horizon function, responsible for the resctriction to the Gribov region corresponding  the the class of the linear covariant gauges, should be constructed only with the transverse component of the gauge field, $A^{T}_{\mu}= (\delta_{\mu\nu} - \frac{\partial_\mu \partial_\nu}{\partial^2} )A_\nu$, ensuring that the operator used to define the Gribov region, $\mathcal{M}^{ab}(A^{T})$,  is Hermitian\footnote{In the Landau gauge this result is immediate since the gauge field is already transverse.}. Finally, more recently \cite{Capri:2015ixa,Capri:2016aqq,Capri:2015nzw}, a novel formulation for the Gribov-Zwanziger framework in the linear covariant gauges exhibiting an exact nonperturbative BRST invariance   was proposed. In this new approach, the horizon function is written in terms of a gauge invariant field  $\mathbf{A}_{\mu}$, which coincides with the transverse component of the gauge field $A^{T}_{\mu}$ only at the lowest-order in the coupling constant $g$. 

The results achieved in \cite{Capri:2015ixa,Capri:2016aqq,Capri:2015nzw} can be summarized by introducing the corresponding Gribov-Zwanziger action in linear covariant gauges as       
\begin{equation}
S^{\mathrm{LCG}}_{\mathrm{GZ}}=S_{\mathrm{FP}}+\gamma^{4}H(\mathbf{A})-4V(N^{2}-1)\gamma^{4}\,, \label{lcgact}
\end{equation}
where $\gamma^{2}$ is the Gribov mass parameter  defined in a self-consistent way through the gap equation: 
\begin{equation}
\langle H (\mathbf{A})\rangle=4V(N^{2}-1)\,,
\end{equation}
and
\begin{equation}
H(\mathbf{A})=g^{2}\int d^{4}xd^{4}y\,f^{abc}\mathbf{A}^{b}_{\mu}(x)\,\left[\mathcal{M}^{-1}(\mathbf{A})\right]^{ad}\!(x,y)\,f^{dec}\mathbf{A}^{e}_{\mu}(y)\,,
\end{equation}
is the horizon function written in terms of the non-polynomial, transverse and gauge invariant field $\mathbf{A}_{\mu}$, given by 
\begin{eqnarray}
\mathbf{A}^{a}_{\mu}=\mathbf{A}^{a}_{\mu}T^{a}&=&e^{-ig\,\xi}\,A_{\mu}\,e^{ig\,\xi}+\frac{i}{g}\,e^{-ig\,\xi}\,\partial_{\mu}e^{ig\,\xi}\,,\nonumber\\
&=&T^{a}\left(A^{a}_{\mu}-D^{ab}_{\mu}\xi^{b}-\frac{g}{2}f^{abc}\xi^{b}\,D^{cd}_{\mu}\xi^{d}+O(\xi^{3})\right).
\label{local_inv_A}
\end{eqnarray}
The auxiliary dimensionless field $\xi=\xi^{a}T^{a}$ is a Stueckelberg-like field, see \cite{Capri:2016aqq}. The gauge invariance of $\mathbf{A}_{\mu}$ can be checked order by order from the transformations
\begin{eqnarray}
\delta A^{a}_{\mu}&=&-D^{ab}_{\mu}\omega^{b}\,,\nonumber\\
\delta \xi^{a}&=&-\left(\delta^{ab}-\frac{g}{2}\,f^{abc}\xi^{c}+\frac{g^{2}}{12}\,f^{acd}f^{cbe}\xi^{d}\xi^{e}+O(\xi^{3})\right)\omega^{b}\,,    \label{gtxi}
\end{eqnarray}
with $\omega^{a}$ being the  infinitesimal parameter of the gauge transformation. From equations \eqref{gtxi} it turns out that $\mathbf{A}^{a}_{\mu}$ is left invariant by the gauge transformations \cite{Capri:2015ixa}, namely  
\begin{equation}
\delta \mathbf{A}^{a}_{\mu} = 0  \;. \label{ginvA}
\end{equation}
In addition, one requires that $\mathbf{A}^{a}_{\mu}$ is transverse, see Appendix A of \cite{Capri:2015ixa} for details, {\it i.e.} 
\begin{equation}
\partial_\mu \mathbf{A}^{a}_{\mu} =0 \;. \label{trA}
\end{equation}

The transversality condition \eqref{trA} allows us to eliminate the auxiliary Stueckelberg field \cite{Capri:2015ixa},  giving  $\mathbf{A}_{\mu}$ as a non-local power series in $A_{\mu}$:
\begin{equation}
\mathbf{A}_{\mu}=\left(\delta_{\mu\nu}-\frac{\partial_{\mu}\partial_{\nu}}{\partial^{2}}\right)\left(A_{\nu}
-ig\left[\frac{\partial\cdot A}{\partial^{2}}\,,\,A_{\nu}\right]
+\frac{ig}{2}\left[\frac{\partial\cdot A}{\partial^{2}}\,,\,\partial_{\nu}\frac{\partial\cdot A}{\partial^{2}}\right]\right)
+O(A^{3})\,.\label{non_local_inv_A}
\end{equation}
Thus, equation \eqref{local_inv_A} is the local version of \eqref{non_local_inv_A}, subject to the transversality constraint $\partial_{\mu}\mathbf{A}_{\mu}=0$. Notice also that, at lowest order, $\mathbf{A}_{\mu}$ coincides with the transverse component of the gauge field, namely
\begin{equation}
\mathbf{A}_{\mu}=\left(\delta_{\mu\nu}-\frac{\partial_{\mu}\partial_{\nu}}{\partial^{2}}\right)A_{\nu}
+O(A^{2})\equiv A^{T}_{\mu}+O(A^{2})\approx\ A^{T}_{\mu}\,.\label{approx}
\end{equation}
Conceptually, the formulation of the Gribov-Zwanziger action in terms of $\mathbf{A}_{\mu}$ brings many non-trivial features \cite{Capri:2015ixa,Capri:2016aqq,Capri:2015nzw}. In particular, it makes possible to introduce a nonperturbative BRST symmetry which enables us to prove the independence from the gauge parameter $\alpha$ of gauge-invariant correlation functions. Also, it allows to establish the non-renormalization of the longitudinal component of the gauge field two-point function \cite{Capri:2016aqq}, in agreement with the results of the most recent lattice numerical simulations \cite{Cucchieri:2009kk,Bicudo:2015rma}. It is worth underlining that  this proposal  \cite{Capri:2015ixa,Capri:2015nzw} has been cast in local form  \cite{Capri:2016aqq}, providing thus a nonperturbative and invariant framework which implements the restriction of the path integral domain to a region free of a large set of Gribov copies in the linear covariant gauges.  

Another relevant feature of the Gribov-Zwanziger approach, not only in the Landau gauge, but also in the maximal Abelian, linear covariant and Coulomb gauges is the dynamical formation of dimension two condensates, see \cite{Capri:2015nzw,Dudal:2008sp,Capri:2015pfa,Guimaraes:2015bra}. The inclusion of such operators in the Gribov-Zwanziger action gives rise to the so-called refined Gribov-Zwanziger action. It is important to emphasize that,  relying on the available lattice data, the gluon propagator computed out of the refined Gribov-Zwanziger action in the aforementioned gauges is in very good agreement with the numerical results. 

We also meantion that, in recent years, several groups started to study the nonperturbative infrared behavior of correlation functions in the linear covariant gauges, see for instance  \cite{Aguilar:2015nqa,Huber:2015ria,Cucchieri:2009kk,Bicudo:2015rma,Siringo:2014lva,Machado:2016cij,Moshin:2015gsa}. The collection of results obtained so far has provided a fruitful interplay among the various approaches available, a strategy which turned out to be very successful in the case of the Landau gauge. 

A question which naturally arises at this stage is whether the framework encoded into the action \eqref{lcgact}  is renormalizable or not.  This is topic which will be faced in the following. In particular, we shall establish the renormalizability to all orders of the action \eqref{lcgact} in the approximation \eqref{approx}, {\it i.e.} in the formulation presented in \cite{Capri:2015pja}.  This first nontrivial step will enable us to set the necessary tools to attack the more complex and difficult problem of the renormalizability of the full action \eqref{lcgact}. 

The paper is organized as follows. In Sect.~2, we present the construction of the local version of the model. In Sect.~3, we display the full set of Ward identities.  In Sect.~4, we prove the renormalizability of model to all orders by means of the algebraic renornalization. In Sect.~5, we discuss the introduction of the dynamical dimension two condensates. Finally,  in Sect~6 we present our conclusions. For completeness, we have included two Appendices containing the tree level propagators of the theory as well as the quantum numbers of all fields and external sources.

\section{The model}
\subsection{A local formulation of the horizon function}\label{horizonlocalization}
As stated in the introduction, we will work explicitly  in the approximation \eqref{approx}. Then, the horizon function we consider here is
\begin{equation}
H(A^{T})=g^{2}\int d^{4}xd^{4}y\,f^{abc}(A^T)^{a}_{\mu}(x)[\mathcal{M}^{-1}(A^T)]^{bd}(x,y)\,f^{edc}(A^T)^{e}_{\mu}(y)\,.
\label{horizon_LCG}
\end{equation}
Therefore, the corresponding Gribov-Zwanziger action in the linear covariant gauges  reads
\begin{equation}
S=S_{\mathrm{FP}}+\gamma^{4}\,H(A^{T})\,,
\label{GZ-LCG}
\end{equation}
which, neglecting for the moment the vacuum term $4V(N^{2}-1)\gamma^{4}$,  coincides with expression \eqref{lcgact} when $\mathbf{A}_{\mu}\approx A^{T}_{\mu}$. Written in this fashion, the Gribov-Zwanziger action in linear covariant gauges \eqref{GZ-LCG} is non-local. In fact, it contains two types of non-localities: first, the transverse component of the gauge field is non-local, as expressed by eq.\eqref{approx}. Second, the horizon function itself $H(A^{T})$ is non-local. 

In the original Gribov-Zwanziger construction in the Landau gauge, a localization procedure for the horizon function has been worked out. Though, in order to employ it in the present case, we  first need to express the transverse component of the gauge field in local form. In order to achieve this task, we replace the action $S$ by

\begin{equation}
S_{\mbox{\scriptsize off-shell}}=S_{\mathrm{FP}}+\gamma^{4}\,H(B)+\int d^{4}x\, \left[\,i\rho^{a}\,\partial_{\mu}B^{a}_{\mu}
+\lambda^{a}_{\mu}\,(B^{a}_{\mu}-(A^{a}_{\mu}-\partial_{\mu}\xi^{a}))\,\right]\,,
\end{equation}
where we have introduced a new set of  fields $\{B^{a}_{\mu},\xi^{a},\rho^{a},\lambda^{a}_{\mu}\}$, with $\rho^{a}$ and $\lambda^{a}_{\mu}$ corresponding to Lagrange multipliers enforcing the following constraints:
\begin{eqnarray}
\frac{\delta S_{\mbox{\scriptsize off-shell}}}{\delta \rho^{a}}=i\partial_{\mu}B^{a}_{\mu}=0\qquad&\Rightarrow&\qquad  B^{a}_{\mu}\equiv(B^T)^{a}_{\mu}=\left(\delta_{\mu\nu}-\frac{\partial_{\mu}\partial_{\nu}}{\partial^{2}}\right)\,B^{a}_{\nu}\,.\label{rho_constraint}\\
\frac{\delta S_{\mbox{\scriptsize off-shell}}}{\delta \lambda^{a}_{\mu}}=B^{a}_{\mu}+\partial_{\mu}\xi^{a}-A^{a}_{\mu}=0\qquad&\Rightarrow&\qquad A^{a}_{\mu}=B^{a}_{\mu}+\partial_{\mu}\xi^{a}\,.\label{lambda_constraint}
\end{eqnarray}
On the other hand, the gauge field can always be decomposed into transverse and longitudinal components:
\begin{equation}
A_{\mu}=A^{T}_{\mu}+A^{L}_{\mu}\,.   \label{dec}
\end{equation}
Combining thus eq.\eqref{dec} with the constraints \eqref{rho_constraint} and \eqref{lambda_constraint}, we get the following on-shell relations
\begin{equation}
B_{\mu}=A^{T}_{\mu}\,,\qquad \partial_{\mu}\xi=A^{L}_{\mu}\,.     \label{B} 
\end{equation}
Integrating out the fields  $\rho^{a}$ and $\lambda^{a}_{\mu}$, we can show that the actions $S$ and $S_{\mbox{\scriptsize off-shell}}$ are equivalent. Thus, $S_{\mbox{\scriptsize off-shell}}$ is the off-shell version of $S$. Now, one can localize the horizon function expressed in terms of the local field $B^{a}_{\mu}$ present in the action $S_{\mbox{\scriptsize off-shell}}$ by the introduction of a set of auxiliary localizing fields $\{\varphi^{ab}_{\mu},\bar\varphi^{ab}_{\mu}, \omega^{ab}_{\mu},\bar\omega^{ab}_{\mu}\}$, namely 
\begin{eqnarray}
S_{\mbox{\scriptsize off-shell}}^{\mathrm{local}}&=&S_{\mathrm{FP}}+\int d^{4}x\,\bigg[\,\bar\varphi^{ac}_{\mu}\,\partial_{\nu}D^{ab}_{\nu}(B)\varphi^{bc}_{\mu}
-\bar\omega^{ac}_{\mu}\,\partial_{\nu}D^{ab}_{\nu}(B)\omega^{bc}_{\mu}
+\gamma^{2}gf^{abc}B^{a}_{\mu}(\varphi^{bc}_{\mu}+\bar\varphi^{bc}_{\mu})\nonumber\\
&&+i\rho^{a}\,\partial_{\mu}B^{a}_{\mu}
+\lambda^{a}_{\mu}\,(B^{a}_{\mu}-(A^{a}_{\mu}-\partial_{\mu}\xi^{a}))\,\bigg]\,,
\label{S_local}
\end{eqnarray}
where $\{\varphi^{ab}_{\mu},\bar\varphi^{ab}_{\mu}\}$ are a pair of  bosonic fields and $\{\omega^{ab}_{\mu},\bar\omega^{ab}_{\mu}\}$ a pair of Grassmannian fields. The integration of such auxiliary fields gives back  the non-local expression in terms of the horizon function $H(A^T)$. Therefore, expression \eqref{S_local} corresponds to a local version of the Gribov-Zwanziger action in linear covariant gauges in the approximation \eqref{approx}.

\subsection{The BRST invariance}

In the last section, we have faced the problem of the localization of the horizon function. In order to prove the renormalizability of \eqref{S_local}  to all orders in perturbation theory,  we will employ the algebraic renormalization set up \cite{Piguet:1995er}. Therefore, an essential tool is the BRST symmetry and its cohomology. As it happens in the Landau gauge in the original Gribov-Zwanziger construction, the action \eqref{S_local} in the approximation \eqref{approx} breaks the BRST symmetry in an explicit way.  Nevertheless, the breaking is soft due to the presence of the Gribov parameter $\gamma$. This problem is circumvented by embedding the theory into an extended  BRST invariant one by the introduction of a suitable set of external sources. We shall proceed thus by casting the action \eqref{S_local} in a BRST invariant fashion. 

First, let us remind that the Faddeev-Popov action, $S_{\mathrm{FP}}$, is left invariant by the usual BRST transformations:
\begin{eqnarray}
sA^{a}_{\mu}&=&-D^{ab}_{\mu}(A)c^{b}\,,\nonumber\\
sc^{a}&=&\frac{g}{2}f^{abc}c^{b}c^{c}\,,\nonumber\\
s\bar{c}^{a}&=&ib^{a}\,,\nonumber\\
sb^{a}&=&0\,.
\end{eqnarray}
Also, it can be easily checked that the BRST operator, $s$, is nilpotent, {\it i.e.}  $s^{2}=0$. Following \cite{Zwanziger:1989mf}, in order to keep the nilpotency of $s$, the remaining auxiliary fields $$\{\varphi^{ab}_{\mu},\varphi^{ab}_{\mu},\omega^{ab}_{\mu},\bar\omega^{ab}_{\mu},B^{a}_{\mu},\xi^{a},\rho^{a},\lambda^{a}_{\mu}\}\,,$$ are required to transform as BRST doublets, namely 
\begin{eqnarray}
&s\varphi^{ab}_{\mu}=\omega^{ab}_{\mu}\,,\qquad s\omega^{ab}_{\mu}=0\,,&\nonumber\\
&s\bar\omega^{ab}_{\mu}=\bar\varphi^{ab}_{\mu}\,,\qquad s\bar\varphi^{ab}_{\mu}=0\,,&\nonumber\\
&sB^{a}_{\mu}=\eta^{a}_{\mu}\,,\qquad s\eta^{a}_{\mu}=0\,,&\nonumber\\
&s\xi^{a}=u^{a}\,,\qquad su^{a}=0\,,&\nonumber\\
&s\vartheta^{a}=i\rho^{a}\,,\qquad s\rho^{a}=0\,,&\nonumber\\
&s\tau^{a}_{\mu}=\lambda^{a}_{\mu}\,,\qquad s\lambda^{a}_{\mu}=0\,,&
\end{eqnarray}
so that 
\begin{equation}
s^2 =0 \;. \label{nnilp}
\end{equation}

Notice that, in order to preserve the BRST doublet structure, we needed to introduce the anti-commuting fields $\{\eta^{a}_{\mu}, u^{a},\vartheta^{a},\tau^{a}\}$.

Performing now the following shift in the field $\omega^{ab}_{\mu}$ with unity Jacobian:
\begin{equation}
\omega^{bc}_{\mu}\to\omega^{bc}_{\mu}+\left(\frac{1}{\partial\cdot D}\right)^{bd}\partial_{\nu}\left(gf^{dmn}\eta^{m}_{\nu}\varphi^{nc}_{\mu}\right)\,,
\end{equation}
the action \eqref{S_local} gets replaced by 
\begin{eqnarray}
S_{\mbox{\scriptsize quasi-inv}}&=&S_{\mathrm{FP}}+s\int d^{4}x\,\bigg[\,\bar\omega^{ac}_{\mu}\,\partial_{\nu}D^{ab}_{\nu}(B)\varphi^{bc}_{\mu}+\vartheta^{a}\,\partial_{\mu}B^{a}_{\mu}
+\tau^{a}_{\mu}(B^{a}_{\mu}-(A^{a}_{\mu}-\partial_{\mu}\xi^{a}))\,\bigg]\nonumber\\
&&+\gamma^{2}\int d^{4}x\, gf^{abc}B^{a}_{\mu}(\varphi^{bc}_{\mu}+\bar\varphi^{bc}_{\mu})\,.
\end{eqnarray}
The action  $S_{\mbox{\scriptsize quasi-inv}}$ is not left invariant by the above  BRST transformations, which are softly broken by the terms proportional to $\gamma^{2}$, namely  
\begin{equation}
sS_{\mbox{\scriptsize quasi-inv}}=\gamma^{2}\int d^{4}x\, gf^{abc}\left[\,\eta^{a}_{\mu}(\varphi^{bc}_{\mu}+\bar\varphi^{bc}_{\mu})
+B^{a}_{\mu}\omega^{bc}_{\mu}\,\right]\,.
\end{equation}
Therefore, in order to restore BRST symmetry, we are led to consider an extended action which reduces to  $S_{\mbox{\scriptsize quasi-inv}}$ when a suitable physical limit is taken.  Such extended action is obtained by introducing a set of external sources forming a pair of BRST doublet, {\it i.e.}
\begin{equation}
sM^{ab}_{\mu\nu}=N^{ab}_{\mu\nu}\,,\qquad sN^{ab}_{\mu\nu}=0\,,\qquad
s\bar{N}^{ab}_{\mu\nu}=\bar{M}^{ab}_{\mu\nu}\,,\qquad s\bar{M}^{ab}_{\mu\nu}=0\,,
\end{equation}
and it is given by
\begin{eqnarray}
 S_{\mathrm{inv}}&=&S_{\mathrm{YM}}+s\int d^{4}x\,\bigg[-i\frac{\alpha}{2}\,\bar{c}^{a}b^{a}
+\bar{c}^{a}\,\partial_{\mu}A^{a}_{\mu}+\bar\omega^{ac}_{\mu}\,\partial_{\nu}D^{ab}_{\nu}(B)\varphi^{bc}_{\mu}
-\bar{N}^{ac}_{\mu\nu}\,D^{ab}_{\mu}(B)\varphi^{bc}_{\nu}\nonumber\\
&&
 -M^{ac}_{\mu\nu}\,D^{ab}_{\mu}(B)\bar{\omega}^{bc}_{\nu}
+\kappa\,\bar{N}^{ab}_{\mu\nu}M^{ab}_{\mu\nu}
+\vartheta^{a}\,\partial_{\mu}B^{a}_{\mu}
+\tau^{a}_{\mu}(B^{a}_{\mu}-(A^{a}_{\mu}-\partial_{\mu}\xi^{a}))\,\bigg]\nonumber\\
&=&\int d^{4}x\,\bigg\{\frac{1}{4}\,F^{a}_{\mu\nu}F^{a}_{\mu\nu}
+\frac{\alpha}{2}\,b^{a}b^{a}+ib^{a}\,\partial_{\mu}A^{a}_{\mu}
+\bar{c}^{a}\,\partial_{\mu}D^{ab}_{\mu}(A)c^{b}
+\bar\varphi^{ac}_{\nu}\,\partial_{\mu}D^{ab}_{\mu}(B)\varphi^{bc}_{\nu}
\nonumber\\
\phantom{\Big|}&&
-\bar\omega^{ac}_{\nu}\,\partial_{\mu}D^{ab}_{\mu}(B)\omega^{bc}_{\nu}
+gf^{abc}(\partial_{\mu}\bar\omega^{ad}_{\nu})\eta^{b}_{\mu}\varphi^{cd}_{\nu}
-\bar{M}^{ac}_{\mu\nu}D^{ab}_{\mu}(B)\varphi^{bc}_{\nu}\nonumber\\
\phantom{\Big|}&&
+\bar{N}^{ac}_{\mu\nu}\left[D^{ab}_{\mu}(B)\omega^{bc}_{\nu}+gf^{abd}\eta^{b}_{\mu}\varphi^{dc}_{\nu}\right]
-{N}^{ac}_{\mu\nu}D^{ab}_{\mu}(B)\bar{\omega}^{bc}_{\nu}
-{M}^{ac}_{\mu\nu}\left[D^{ab}_{\mu}(B)\bar\varphi^{bc}_{\nu}+gf^{abd}\eta^{b}_{\mu}\bar\omega^{dc}_{\nu}\right]
\nonumber\\
\phantom{\Big|}&&
+\kappa\,\left(\bar{M}^{ab}_{\mu\nu}M^{ab}_{\mu\nu} - \bar{N}^{ab}_{\mu\nu}N^{ab}_{\mu\nu}\right)
+i\rho^{a}\,\partial_{\mu}B^{a}_{\mu}
-\vartheta^{a}\,\partial_{\mu}\eta^{a}_{\mu}
+\lambda^{a}_{\mu}(B^{a}_{\mu}-(A^{a}_{\mu}-\partial_{\mu}\xi^{a}))\nonumber\\
&&-\tau^{a}_{\mu}(\eta^{a}_{\mu}+\partial_{\mu}u^{a}+D^{ab}_{\mu}(A)c^{b})
\bigg\}\,,
\end{eqnarray}
with 
\begin{equation}
s S_{\mathrm{inv}} = 0 \;. \label{vv}
\end{equation}
Here, we have followed the original Zwanziger approach \cite{Zwanziger:1989mf} in which the external sources $\{M,\bar{M},N,\bar{N}\}$ are introduced  in such a way that when they acquire their physical values
\begin{equation}
M^{ab}_{\mu\nu}\Big|_{phys}=
\bar{M}^{ab}_{\mu\nu}\Big|_{phys}=\gamma^{2}\,\delta^{ab}\delta_{\mu\nu}\,,\qquad
N^{ab}_{\mu\nu}\Big|_{phys}=
\bar{N}^{ab}_{\mu\nu}\Big|_{phys}=0\,,   \label{plm}
\end{equation}
the action $S_{\mathrm{inv}}$ coincides with $S_{\mbox{\scriptsize quasi-inv}}$, {\it i.e.}
\begin{equation}
S_{\mathrm{inv}}\Big|_{phys} = S_{\mbox{\scriptsize quasi-inv}}    \;, \label{pl}
\end{equation}
from which it becomes apparent that $S_{\mathrm{inv}}$ is a BRST  invariant extension of $S_{\mbox{\scriptsize quasi-inv}}$. In particular, renormalizability of $S_{\mathrm{inv}}$ will imply that of $S_{\mbox{\scriptsize quasi-inv}}$. 

We have also added to $S_{\mathrm{inv}}$ the term $\kappa\,(\bar{M}M-\bar{N}N)$, with $\kappa$ being a dimensionless parameter. This term is allowed by power-counting and, in the physical limit \eqref{plm}, represents the vacuum term\footnote{Later will be shown that $\kappa=-1$ and this vaccum term coincides with that in Eq.~\eqref{lcgact}, as expected.}, {\it i.e.}
\begin{equation}
(\bar{M}M-\bar{N}N)\Big|_{phys} = 4 (N^2-1) \gamma^4 \;. 
\end{equation} 
For further use, following the algebraic renornalization procedure \cite{Piguet:1995er}, one has to introduce BRST invariant sources $(\Omega^a_\mu, L^a)$ coupled to the nonlinear BRST teansformations of the fields $(A^a_\mu, c^a)$ 
\begin{equation}
\Sigma_0=S_{\mathrm{inv}}+\int d^{4}x\,\left[\,\Omega^{a}_{\mu}\,(sA^{a}_{\mu})+L^{a}\,(sc^{a})\,\right]\,,  
\end{equation} 
with 
\begin{equation}
s \Omega^a_\mu = 0 \;, \qquad s L^a = 0 \;, 
\end{equation} 
and 
\begin{equation}
s \Sigma_0 = 0 \;. 
\end{equation}
In addition, as shown originally in \cite{Zwanziger:1989mf}, the extended action $\Sigma_0$ allows the introduction of the very useful multi-index notation. In fact, following \cite{Zwanziger:1989mf}, one introduces the composite index:  
\begin{eqnarray}
i&\equiv&(a,\mu)=1,\dots,4(N^{2}-1)\,,\nonumber\\
(\varphi,\bar\varphi,\omega,\bar\omega)^{ab}_{\mu}&\equiv&(\varphi,\bar\varphi,\omega,\bar\omega)^{a}_{i}\,,\nonumber\\
(M,\bar{M},N,\bar{N})^{ab}_{\mu\nu}&\equiv&(M,\bar{M},N,\bar{N})^{a}_{\mu i}\,.
\end{eqnarray}
Therefore, using this notation, the action $\Sigma_0$ can be written as
\begin{eqnarray}
\Sigma_{0}&=&\int d^{4}x\,\bigg\{\frac{1}{4}\,F^{a}_{\mu\nu}F^{a}_{\mu\nu}
+\frac{\alpha}{2}\,b^{a}b^{a}+ib^{a}\,\partial_{\mu}A^{a}_{\mu}
+\bar{c}^{a}\,\partial_{\mu}D^{ab}_{\mu}(A)c^{b}
+\bar\varphi^{a}_{i}\,\partial_{\mu}D^{ab}_{\mu}(B)\varphi^{b}_{i}
-\bar\omega^{a}_{i}\,\partial_{\mu}D^{ab}_{\mu}(B)\omega^{b}_{i}\nonumber\\
\phantom{\Big|}&&
+gf^{abc}(\partial_{\mu}\bar\omega^{a}_{i})\eta^{b}_{\mu}\varphi^{c}_{i}
-\bar{M}^{a}_{\mu i}D^{ab}_{\mu}(B)\varphi^{b}_{i}
+\bar{N}^{a}_{\mu i}\left[D^{ab}_{\mu}(B)\omega^{b}_{i}+gf^{abc}\eta^{b}_{\mu}\varphi^{c}_{i}\right]
-{N}^{a}_{\mu i}D^{ab}_{\mu}(B)\bar{\omega}^{b}_{i}\nonumber\\
\phantom{\Big|}&&
-{M}^{a}_{\mu i}\left[D^{ab}_{\mu}(B)\bar\varphi^{b}_{i}+gf^{abc}\eta^{b}_{\mu}\bar\omega^{c}_{i}\right]
+\kappa\,\left(\bar{M}^{a}_{\mu i}M^{a}_{\mu i} - \bar{N}^{a}_{\mu i}N^{a}_{\mu i}\right)
+i\rho^{a}\,\partial_{\mu}B^{a}_{\mu}
-\vartheta^{a}\,\partial_{\mu}\eta^{a}_{\mu}\nonumber\\
\phantom{\Big|}&&
+\lambda^{a}_{\mu}(B^{a}_{\mu}-(A^{a}_{\mu}-\partial_{\mu}\xi^{a}))
-\tau^{a}_{\mu}(\eta^{a}_{\mu}+\partial_{\mu}u^{a}+D^{ab}_{\mu}(A)c^{b})
-\Omega^{a}_{\mu}\,D^{ab}_{\mu}(A)c^{b}+\frac{g}{2}f^{abc}L^{a}c^{b}c^{c}
\bigg\}\,.
\label{S_inv}
\end{eqnarray}

\subsection{Introducing more constraints}

Let us proceed by taking a close look at the term of the action $\Sigma_{0}$  corresponding to the localization of the horizon function, namely 
\begin{eqnarray}
S_{H}&=&S_{H}[B,\eta,\vartheta,\rho,\varphi,\bar\varphi,\omega,\bar\omega,M,\bar{M},N,\bar{N}]\nonumber\\
&=&\int d^{4}x\,\bigg\{\bar\varphi^{a}_{i}\,\partial_{\mu}D^{ab}_{\mu}(B)\varphi^{b}_{i}
-\bar\omega^{a}_{i}\,\partial_{\mu}D^{ab}_{\mu}(B)\omega^{b}_{i}
+gf^{abc}(\partial_{\mu}\bar\omega^{a}_{i})\eta^{b}_{\mu}\varphi^{c}_{i}
-\bar{M}^{a}_{\mu i}D^{ab}_{\mu}(B)\varphi^{b}_{i}\nonumber\\
\phantom{\Big|}&&
+\bar{N}^{a}_{\mu i}\left[D^{ab}_{\mu}(B)\omega^{b}_{i}+gf^{abc}\eta^{b}_{\mu}\varphi^{c}_{i}\right]
-{N}^{a}_{\mu i}D^{ab}_{\mu}(B)\bar{\omega}^{b}_{i}
-{M}^{a}_{\mu i}\left[D^{ab}_{\mu}(B)\bar\varphi^{b}_{i}+gf^{abc}\eta^{b}_{\mu}\bar\omega^{c}_{i}\right]\nonumber\\
\phantom{\Big|}&&
+\kappa\,\left(\bar{M}^{a}_{\mu i}M^{a}_{\mu i} - \bar{N}^{a}_{\mu i}N^{a}_{\mu i}\right)
+i\rho^{a}\,\partial_{\mu}B^{a}_{\mu}
-\vartheta^{a}\,\partial_{\mu}\eta^{a}_{\mu}\bigg\}\,.
\end{eqnarray}
This term enjoys several useful exact symmetries.  In particular, we  call attention to the following invariances:
\begin{eqnarray}
\mathcal{W}^{a}(S_{H})&=&0\,,\label{W_symm}\\
\overline{\mathcal{W}}^{a}(S_{H})&=&0\,,\label{Wbar_symm}
\end{eqnarray}
where
\begin{eqnarray}
\mathcal{W}^{a}&=&f^{abc}\sum_{y\in\mathcal{Y}}\int d^{4}x\,\,y^{b}(x)\cdot\frac{\delta}{\delta y^{c}(x)}\,,\label{W_op}\\
\overline{\mathcal{W}}^{a}&=&f^{abc}\int d^{4}x\,\bigg(B^{b}_{\mu}\frac{\delta}{\delta\eta^{c}_{\mu}}
-i\vartheta^{b}\frac{\delta}{\delta\rho^{c}}
+\bar\omega^{b}_{i}\frac{\delta}{\delta\bar\varphi^{c}_{i}}
+\varphi^{b}_{i}\frac{\delta}{\delta\omega^{c}_{i}}
+\bar{N}^{b}_{\mu i}\frac{\delta}{\delta\bar{M}^{c}_{\mu i}}
+M^{b}_{\mu i}\frac{\delta}{\delta N^{c}_{\mu i}}\bigg)\,,\label{Wbar_op}
\end{eqnarray}
with
\begin{equation}
\mathcal{Y}=\Big\{B^{a}_{\mu},\eta^{a}_{\mu},\rho^{a},\vartheta^{a},\varphi^{a}_{i},\bar\varphi^{a}_{i},\omega^{a}_{i},\bar\omega^{a}_{i},M^{a}_{\mu i},
\bar{M}^{a}_{\mu i},N^{a}_{\mu i},\bar{N}^{a}_{\mu i}\Big\}\,.
\end{equation}
The symmetry \eqref{W_symm} is recognized to express the invariance under rigid gauge transformations.  As such,  there are no difficulties in generalizing the operator \eqref{W_op} in order to include the remaining fields of the action \eqref{S_inv}, so that the symmetry \eqref{W_symm}   extends to the whole action $\Sigma_{0}$.  Though, this is not the case of the second invariance \eqref{Wbar_symm}, which cannot be extended immediately to the whole action $\Sigma_{0}$, due to the presence  of the terms involving $\lambda^{a}_{\mu}$ and $\tau^{a}_{\mu}$ which give rise to explicit breaking terms. Nevertheless, it turns out to be possible to extend equation \eqref{Wbar_symm} to be an exact symmetry of the model by introducing the following set of BRST doublets of fields:
\begin{equation}
sX^{ab}=Y^{ab}\,,\qquad sY^{ab}=0\,;\qquad
s\widetilde{X}^{ab}=\widetilde{Y}^{ab}\,,\qquad s\widetilde{Y}^{ab}=0\,;
\label{XY}
\end{equation}
\begin{equation}
sT^{ab}=H^{ab}\,,\qquad sH^{ab}=0\,;\qquad
s\widetilde{T}^{ab}=\widetilde{H}^{ab}\,,\qquad s\widetilde{H}^{ab}=0\,;
\label{TH}
\end{equation}
and perform the replacement:
\begin{eqnarray}
s[\tau^{a}_{\mu}(B^{a}_{\mu}-(A^{a}_{\mu}-\partial_{\mu}\xi^{a}))]&\to&s[\tau^{b}_{\mu}(X^{ab}B^{a}_{\mu}-\widetilde{X}^{ab}(A^{a}_{\mu}-\partial_{\mu}\xi^{a}))
+T^{ab}(X^{ab}-\delta^{ab})+\widetilde{T}^{ab}(\widetilde{X}^{ab}-\delta^{ab})]\nonumber\\
&=&\lambda^{b}_{\mu}\,[X^{ab}B^{a}_{\mu}-\widetilde{X}^{ab}(A^{a}_{\mu}-\partial_{\mu}\xi^{a})]-\tau^{b}_{\mu}\,[Y^{ab}B^{a}_{\mu}+X^{ab}\eta^{a}_{\mu}-\widetilde{Y}^{ab}(A^{a}_{\mu}-\partial_{\mu}\xi^{a})\nonumber\\
&&+\widetilde{X}^{ab}(D^{ac}_{\mu}(A)c^{c}+\partial_{\mu}u^{a})]+H^{ab}(X^{ab}-\delta^{ab})+\widetilde{H}^{ab}(\widetilde{X}^{ab}-\delta^{ab})\nonumber\\
&&-T^{ab}Y^{ab}-\widetilde{T}^{ab}\widetilde{Y}^{ab}\,.
\end{eqnarray}
Therefore, from now on, we shall consider the new action  given by
\begin{eqnarray}
\Sigma&=&\int d^{4}x\,\bigg\{\frac{1}{4}\,F^{a}_{\mu\nu}F^{a}_{\mu\nu}
+\frac{\alpha}{2}\,b^{a}b^{a}+ib^{a}\,\partial_{\mu}A^{a}_{\mu}
+\bar{c}^{a}\,\partial_{\mu}D^{ab}_{\mu}(A)c^{b}
+\bar\varphi^{a}_{i}\,\partial_{\mu}D^{ab}_{\mu}(B)\varphi^{b}_{i}
-\bar\omega^{a}_{i}\,\partial_{\mu}D^{ab}_{\mu}(B)\omega^{b}_{i}\nonumber\\
\phantom{\Big|}&&
+gf^{abc}(\partial_{\mu}\bar\omega^{a}_{i})\eta^{b}_{\mu}\varphi^{c}_{i}
-\bar{M}^{a}_{\mu i}D^{ab}_{\mu}(B)\varphi^{b}_{i}
+\bar{N}^{a}_{\mu i}\left[D^{ab}_{\mu}(B)\omega^{b}_{i}+gf^{abc}\eta^{b}_{\mu}\varphi^{c}_{i}\right]
-{N}^{a}_{\mu i}D^{ab}_{\mu}(B)\bar\omega^{b}_{i}\nonumber\\
\phantom{\Big|}&&
-{M}^{a}_{\mu i}\left[D^{ab}_{\mu}(B)\bar\varphi^{b}_{i}+gf^{abc}\eta^{b}_{\mu}\bar\omega^{c}_{i}\right]
+\kappa\,\left(\bar{M}^{a}_{\mu i}M^{a}_{\mu i} - \bar{N}^{a}_{\mu i}N^{a}_{\mu i}\right)
+i\rho^{a}\,\partial_{\mu}B^{a}_{\mu}
-\vartheta^{a}\,\partial_{\mu}\eta^{a}_{\mu}\nonumber\\
\phantom{\Big|}&&
+\lambda^{b}_{\mu}\,\Big[X^{ab}B^{a}_{\mu}-\widetilde{X}^{ab}(A^{a}_{\mu}-\partial_{\mu}\xi^{a})\Big]
-\tau^{b}_{\mu}\,\Big[Y^{ab}B^{a}_{\mu}+X^{ab}\eta^{a}_{\mu}-\widetilde{Y}^{ab}(A^{a}_{\mu}-\partial_{\mu}\xi^{a})\nonumber\\
&&
+\widetilde{X}^{ab}(D^{ac}_{\mu}(A)c^{c}+\partial_{\mu}u^{a})\Big]+H^{ab}(X^{ab}-\delta^{ab})+\widetilde{H}^{ab}(\widetilde{X}^{ab}-\delta^{ab})
-T^{ab}Y^{ab}-\widetilde{T}^{ab}\widetilde{Y}^{ab}\nonumber\\
&&
-\Omega^{a}_{\mu}\,D^{ab}_{\mu}(A)c^{b}
+\frac{g}{2}f^{abc}L^{a}c^{b}c^{c}\bigg\}\,.\label{action}
\end{eqnarray}
From this expression one sees that the new fields $(H^{ab}, \widetilde{H}^{ab})$ yield the following exact linearly broken Ward identities 
\begin{eqnarray}
\frac{\delta\Sigma}{\delta H^{ab}}=X^{ab}-\delta^{ab}  \,;\label{H_constraint_a1}\\
\frac{\delta\Sigma}{\delta \widetilde{H}^{ab}}=\widetilde{X}^{ab}-\delta^{ab} \,. \label{Htilde_constraint_a2}
\end{eqnarray} 
The right hand sides of eqs.\eqref{H_constraint_a1},\eqref{Htilde_constraint_a2} are linear in the quantum fields, so that  these terms are linear breaking,  not affected by the quantum corrections \cite{Piguet:1995er}.  The physical meaning expressed by eqs.\eqref{H_constraint_a1},\eqref{Htilde_constraint_a2} is well captured by looking at the equation of motion of the field $\lambda^a_\mu$, {\it i.e.}  
\begin{equation}
\frac{\delta\Sigma}{\delta\lambda^{a}_{\mu}}=X^{ba}B^{b}_{\mu}-\widetilde{X}^{ba}(A^{b}_{\mu}-\partial_{\mu}\xi^{b})=0 \;, \label{ac1}
\end{equation}
which, due to \eqref{H_constraint_a1},\eqref{Htilde_constraint_a2}, gives 
\begin{eqnarray}
 X^{ab}=\delta^{ab}\,; \qquad  \widetilde{X}^{ab}=\delta^{ab}\,;\label{Htilde_constraint}\\
X^{ba}B^{b}_{\mu}-\widetilde{X}^{ba}(A^{b}_{\mu}-\partial_{\mu}\xi^{b})=0\qquad&\Rightarrow&\qquad A^{a}_{\mu}=B^{a}+\partial_{\mu}\xi^{a}\,,\label{new_lambda_constraint}
\end{eqnarray}
so that expression \eqref{lambda_constraint}  is recovered. Besides the two linearly broken identities \eqref{H_constraint_a1},\eqref{Htilde_constraint_a2}, the action $\Sigma$ enjoys the following additional Ward identity, also linearly broken:
\begin{equation}
\mathcal{Q}^{ab}(\Sigma)=-H^{ba}-\widetilde{H}^{ba}\,,\label{Q_symm}
\end{equation}
where
\begin{eqnarray}
\mathcal{Q}^{ab}&=&\lambda^{a}_{\mu}\frac{\delta}{\delta\lambda^{b}_{\mu}}
+\tau^{a}_{\mu}\frac{\delta}{\delta\tau^{b}_{\mu}}
+H^{ca}\frac{\delta}{\delta H^{cb}}
+\widetilde{H}^{ca}\frac{\delta}{\delta \widetilde{H}^{cb}}
+T^{ca}\frac{\delta}{\delta T^{cb}}
+\widetilde{T}^{ca}\frac{\delta}{\delta \widetilde{T}^{cb}}\nonumber\\
&-& X^{cb}\frac{\delta}{\delta X^{ca}}
-\widetilde{X}^{cb}\frac{\delta}{\delta \widetilde{X}^{ca}}
-Y^{cb}\frac{\delta}{\delta Y^{ca}}
-\widetilde{Y}^{cb}\frac{\delta}{\delta \widetilde{Y}^{ca}}\,.
\end{eqnarray}
Moreover, taking the trace of $\mathcal{Q}^{ab}$ in color space, one gets the charge $\mathcal{Q}$ 
\begin{eqnarray}
\mathrm{tr}[\mathcal{Q}^{ab}(x)]\equiv\mathcal{Q}&=&\int d^{4}x\,\bigg(\lambda^{a}_{\mu}\frac{\delta}{\delta\lambda^{a}_{\mu}}
+\tau^{a}_{\mu}\frac{\delta}{\delta\tau^{a}_{\mu}}
+H^{ab}\frac{\delta}{\delta H^{ab}}
+\widetilde{H}^{ab}\frac{\delta}{\delta \widetilde{H}^{ab}}
+T^{ab}\frac{\delta}{\delta T^{ab}}
+\widetilde{T}^{ab}\frac{\delta}{\delta \widetilde{T}^{ab}}\nonumber\\
&-&X^{ab}\frac{\delta}{\delta X^{ab}}
-\widetilde{X}^{ab}\frac{\delta}{\delta \widetilde{X}^{ab}}
-Y^{ab}\frac{\delta}{\delta Y^{ab}}
-\widetilde{Y}^{ab}\frac{\delta}{\delta \widetilde{Y}^{ab}}\bigg)\,.    \label{charge}
\end{eqnarray}
As one sees from eq.\eqref{charge}, for the fields $\{\lambda,\tau,H,\widetilde{H},T,\widetilde{T}\}$, the value  of the  charge $\mathcal{Q}$ is $+1$, while for the fields $\{X,\widetilde{X},Y,\widetilde{Y}\}$  is $-1$. As a consequence of the charge assignment, it follows that the dimensionless fields  $\{X,\widetilde{X},Y,\widetilde{Y}\}$ appear in the counterterm always in combination with the dimension 3 fields $\{\lambda,\tau,H,\widetilde{H},T,\widetilde{T}\}$ in order to produce a term with zero charge $\mathcal{Q}$. We see therefore that the charge $\mathcal{Q}$ is very useful in order to keep control of the dependence of the  local invariant counterterm from the fields $\{X,\widetilde{X},Y,\widetilde{Y}\}$.   Furthermore, the equations of motion of the fields $\{\lambda,\tau,H,\widetilde{H},T,\widetilde{T}\}$ correspond to Ward identities, as it will be shown in the next section.

Finally, let us mention that, with the introduction of the BRST doublets \eqref{XY},\eqref{TH}, the equations \eqref{W_symm} and \eqref{Wbar_symm}  can be promoted to Ward identities for  the complete action $\Sigma$, forbidding, in particular, the mixing between the fields $B^{a}_{\mu}$ and $A^{a}_{\mu}$. We will come back to the analysis of such symmetries in the next section.  

\section{Symmetries and Ward identities}

As mentioned before, the action $\Sigma$ obeys a large set of Ward identities, which we enlist below: 
\begin{itemize}
\item{The Slavnov-Taylor identity:
\begin{eqnarray}
\mathcal{S}(\Sigma)&=&\int d^{4}x\,\bigg(
\frac{\delta\Sigma}{\delta\Omega^{a}_{\mu}}\frac{\delta\Sigma}{\delta A^{a}_{\mu}}
+\frac{\delta\Sigma}{\delta L^{a}}\frac{\delta\Sigma}{\delta c^{a}}
+i{b}^{a}\frac{\delta\Sigma}{\delta\bar{c}^{a}}
+\eta^{a}_{\mu}\frac{\delta\Sigma}{\delta B^{a}_{\mu}}
+i\rho^{a}\frac{\delta\Sigma}{\delta \vartheta^{a}}
+u^{a}\frac{\delta\Sigma}{\delta \xi^{a}}\nonumber\\
&+&\omega^{a}_{i}\frac{\delta\Sigma}{\delta \varphi^{a}_{i}}
+\bar\varphi^{a}_{i}\frac{\delta\Sigma}{\delta\bar\omega^{a}_{i}}
+N^{a}_{\mu i}\frac{\delta\Sigma}{\delta M^{a}_{\mu i}}
+\bar{M}^{a}_{\mu i}\frac{\delta\Sigma}{\delta\bar{N}^{a}_{\mu i}}
+\lambda^{a}_{\mu}\frac{\delta\Sigma}{\delta\tau^{a}_{\mu}}
+Y^{ab}\frac{\delta\Sigma}{\delta X^{ab}}\nonumber\\
&+&\widetilde{Y}^{ab}\frac{\delta\Sigma}{\delta \widetilde{X}^{ab}}
+H^{ab}\frac{\delta\Sigma}{\delta T^{ab}}
+\widetilde{H}^{ab}\frac{\delta\Sigma}{\delta \widetilde{T}^{ab}}\,\bigg)\nonumber\\
&=&0\,.\label{ST}
\end{eqnarray}
}
\item{The gauge-fixing and the antighost equations:
\begin{equation}
\frac{\delta\Sigma}{\delta b^{a}}=\alpha b^{a}+i\partial_{\mu}A^{a}_{\mu}\,,\qquad
\frac{\delta\Sigma}{\delta \bar{c}^{a}}+\partial_{\mu}\frac{\delta\Sigma}{\delta \Omega^{a}_{\mu}}=0\,.
\end{equation}
}
\item{The equations of motion of the fields $(\rho^a, \vartheta^a, H^{ab}, \widetilde{H}^{ab}, T^{ab}, \widetilde{T}^{ab})$ and the parametric equation with respect to the parameter $\kappa$:
\begin{eqnarray}
\frac{\delta\Sigma}{\delta \rho^{a}}&=&i\partial_{\mu}B^{a}_{\mu}\,,\\
\frac{\delta\Sigma}{\delta \vartheta^{a}}&=&-\partial_{\mu}\eta^{a}_{\mu}\,,\\
\frac{\delta\Sigma}{\delta H^{ab}}&=&X^{ab}-\delta^{ab}\,,\\
\frac{\delta\Sigma}{\delta \widetilde{H}^{ab}}&=&\widetilde{X}^{ab}-\delta^{ab}\,,\\
\frac{\delta\Sigma}{\delta T^{ab}}&=& -Y^{ab}\,,\\
\frac{\delta\Sigma}{\delta \widetilde{T}^{ab}}&=& -\widetilde{Y}^{ab}\,\\
\frac{\partial\Sigma}{\partial \kappa}&=& \int d^{4}x\,\Big(\bar{M}^{a}_{\mu i}M^{a}_{\mu i}-\bar{N}^{a}_{\mu i}N^{a}_{\mu i}\Big)\,.
\end{eqnarray}
}
\item{Equations of motion of the fields $\lambda$, $\tau$, $\xi$ and $u$:
\begin{equation}
\frac{\delta\Sigma}{\delta\lambda^{a}_{\mu}}
-B^{b}_{\mu}\frac{\delta\Sigma}{\delta H^{ba}}
+(A^{b}_{\mu}-\partial_{\mu}\xi^{b})\frac{\delta\Sigma}{\delta \widetilde{H}^{ba}}=B^{a}_{\mu}-(A^{a}_{\mu}-\partial_{\mu}\xi^{a})\,,
\end{equation}
\begin{equation}
\frac{\delta\Sigma}{\delta\tau^{a}_{\mu}}
-B^{b}_{\mu}\frac{\delta\Sigma}{\delta T^{ba}}
+\eta^{b}_{\mu}\frac{\delta\Sigma}{\delta H^{ba}}
+(A^{b}_{\mu}-\partial_{\mu}\xi^{b})\frac{\delta\Sigma}{\delta\widetilde{T}^{ba}}
+(\partial_{\mu}u^{b})\frac{\delta\Sigma}{\delta\widetilde{H}^{ba}}
-\frac{\delta\Sigma}{\delta\Omega^{a}_{\mu}}
-\frac{\delta\Sigma}{\delta\Omega^{b}_{\mu}}\frac{\delta\Sigma}{\delta\widetilde{H}^{ba}}=-\eta^{a}_{\mu}-(\partial_{\mu}u^{a})\,,
\end{equation}
\begin{equation}
\frac{\delta\Sigma}{\delta \xi^{a}}
+\partial_{\mu}\left(\lambda^{b}_{\mu}\frac{\delta\Sigma}{\delta\widetilde{H}^{ab}}
+\tau^{b}_{\mu}\frac{\delta\Sigma}{\delta\widetilde{T}^{ab}}\right)=- \partial_{\mu}\lambda^{a}_{\mu}\,,
\end{equation}
\begin{equation}
\frac{\delta\Sigma}{\delta u^{a}}+\partial_{\mu}\left(\tau^{b}_\mu\frac{\delta\Sigma}{\delta\widetilde{H}^{ab}}\right)
=-\partial_{\mu}\tau^{a}_\mu\,.
\end{equation}
}
\item{The full rigid symmetry:
\begin{equation}
\mathcal{W}^{a}_{Rigid}(\Sigma)=0\,,
\end{equation}
where
\begin{eqnarray}
\mathcal{W}^{a}_{Rigid}&=&f^{abc}\sum_{y\in \mathcal{G}}\int d^{4}x\, y^{b}(x)\cdot\frac{\delta}{\delta y^{c}(x)}\nonumber\\
&&+f^{abc}\sum_{y\in\mathcal{F}}\int d^{4}x\,\left(y^{bd}(x)\frac{\delta}{\delta y^{cd}(x)}+y^{db}(x)\frac{\delta}{\delta y^{dc}(x)}\right)\,,
\end{eqnarray}
with
\begin{equation}
\mathcal{G}=\mathcal{Y}\,\cup\,\Big\{A^{a}_{\mu},b^{a},c^{a},\bar{c}^{a},\xi^{a},u^{a},L^{a},\Omega^{a}_{\mu},\lambda^{a}_{\mu},
\tau^{a}_{\mu}\Big\}
\end{equation}
and
\begin{equation}
\mathcal{F}=\Big\{X^{ab},\widetilde{X}^{ab},Y^{ab},\widetilde{Y}^{ab},H^{ab},\widetilde{H}^{ab},T^{ab},\widetilde{T}^{ab}\Big\}\,.
\end{equation}
}
\item{The linearly broken Ward identities:
\begin{equation}
\mathcal{W}^{a}(\Sigma)=f^{abc}\int d^{4}x\,H^{bc}\,,
\label{W_symmetry}
\end{equation}
\begin{eqnarray}
\mathcal{W}^{a}&=&f^{abc}\int d^{4}x\,\bigg(\,B^{b}_{\mu}\frac{\delta}{\delta B^{c}_{\mu}}
+\eta^{b}_{\mu}\frac{\delta}{\delta \eta^{c}_{\mu}}
+\rho^{b}\frac{\delta}{\delta \rho^{c}}
+\vartheta^{b}\frac{\delta}{\delta\vartheta ^{c}}
+\varphi^{b}_{i}\frac{\delta}{\delta \varphi^{c}_{i}}
+\bar\varphi^{b}_{i}\frac{\delta}{\delta \bar\varphi^{c}_{i}}
+\omega^{b}_{i}\frac{\delta}{\delta \omega^{c}_{i}}\nonumber\\
&&
+\bar\omega^{b}_{i}\frac{\delta}{\delta \bar\omega^{c}_{i}}
+M^{b}_{\mu i}\frac{\delta}{\delta M^{c}_{\mu i}}
+\bar{M}^{b}_{\mu i}\frac{\delta}{\delta \bar{M}^{c}_{\mu i}}
+N^{b}_{\mu i}\frac{\delta}{\delta N^{c}_{\mu i}}
+\bar{N}^{b}_{\mu i}\frac{\delta}{\delta \bar{N}^{c}_{\mu i}}
+X^{bd}\frac{\delta}{\delta X^{cd}}
\nonumber\\
&&
+Y^{bd}\frac{\delta}{\delta Y^{cd}}
+H^{bd}\frac{\delta}{\delta H^{cd}}
+T^{bd}\frac{\delta}{\delta T^{cd}}
\,\bigg)\,,
\end{eqnarray}
as well as 
\begin{equation}
\overline{\mathcal{W}}^{a}(\Sigma)=-f^{abc}\int d^{4}x\,T^{bc}\,,
\label{Wbar_symmetry}
\end{equation}
\begin{eqnarray}
\overline{\mathcal{W}}^{a}&=&f^{abc}\int d^{4}x\,\bigg(\,B^{b}_{\mu}\frac{\delta}{\delta \eta^{c}_{\mu}}
-i\vartheta^{b}\frac{\delta}{\delta\rho ^{c}}
+\varphi^{b}_{i}\frac{\delta}{\delta \omega^{c}_{i}}
+\bar\omega^{b}_{i}\frac{\delta}{\delta \bar\varphi^{c}_{i}}
+M^{b}_{\mu i}\frac{\delta}{\delta N^{c}_{\mu i}}
+\bar{N}^{b}_{\mu i}\frac{\delta}{\delta \bar{M}^{c}_{\mu i}}\nonumber\\
&&
+X^{bd}\frac{\delta}{\delta Y^{cd}}
+T^{bd}\frac{\delta}{\delta H^{cd}}
\,\bigg)\,.
\end{eqnarray}
}
\item{Equations of motion of the localizing Zwanziger fields:
\begin{eqnarray}
\frac{\delta\Sigma}{\delta\bar\varphi^{a}_{i}}+\partial_{\mu}\frac{\delta\Sigma}{\delta\bar{M}^{a}_{\mu i}}&\!\!\!=\!\!\!&(1+\kappa)\partial_{\mu}M^{a}_{\mu i}-gf^{abc}M^{b}_{\mu i}B^{c}_{\mu}\,,\label{eq_motion_1}\\
\frac{\delta\Sigma}{\delta\bar\omega^{a}_{i}}+\partial_{\mu}\frac{\delta\Sigma}{\delta\bar{N}^{a}_{\mu i}}&=&-(1+\kappa)\partial_{\mu}N^{a}_{\mu i}+gf^{abc}N^{b}_{\mu i}B^{c}_{\mu}\nonumber\\
&&+gf^{abc}M^{b}_{\mu i}\eta^{c}_{\mu}\,,\\
\frac{\delta\Sigma}{\delta\omega^{a}_{i}}+\partial_{\mu}\frac{\delta\Sigma}{\delta{N}^{a}_{\mu i}}+igf^{abc}\bar\omega^{b}_{i}\frac{\delta\Sigma}{\delta\rho^{c}}
&\!\!\!=\!\!\!&(1+\kappa)\partial_{\mu}\bar{N}^{a}_{\mu i}-gf^{abc}\bar{N}^{b}_{\mu i}B^{c}_{\mu}\,,\\
\!\!\!\!\!\!\frac{\delta\Sigma}{\delta\varphi^{a}_{i}}+\partial_{\mu}\frac{\delta\Sigma}{\delta{M}^{a}_{\mu i}}+igf^{abc}\bar\varphi^{b}_{i}\frac{\delta\Sigma}{\delta\rho^{c}}
-gf^{abc}\bar\omega^{b}_{i}\frac{\delta\Sigma}{\delta\vartheta^{c}}
&\!\!\!=\!\!\!&(1+\kappa)\partial_{\mu}\bar{M}^{a}_{\mu i}-gf^{abc}\bar{M}^{b}_{\mu i}B^{c}_{\mu}\nonumber\\
&&+gf^{abc}\bar{N}^{b}_{\mu i}\eta^{c}_{\mu}\,.
\label{eq_motion_4}
\end{eqnarray}
}
\item{The $\mathcal{Q}_{ij}$, $\mathcal{Q}^{ab}$ and ghost number Ward identities:
\begin{eqnarray}
\mathcal{Q}_{ij}(\Sigma)&=&\int d^{4}x\,\bigg(
\varphi^{a}_{i}\frac{\delta\Sigma}{\delta\varphi^{a}_{j}}
-\bar\varphi^{a}_{j}\frac{\delta\Sigma}{\delta\bar\varphi^{a}_{i}}
+\omega^{a}_{i}\frac{\delta\Sigma}{\delta\omega^{a}_{j}}
-\bar\omega^{a}_{j}\frac{\delta\Sigma}{\delta\bar\omega^{a}_{i}}
+M^{a}_{\mu i}\frac{\delta\Sigma}{\delta M^{a}_{\mu j}}
-\bar{M}^{a}_{\mu j}\frac{\delta\Sigma}{\delta\bar{M}^{a}_{\mu i}}\nonumber\\
&&
+N^{a}_{\mu i}\frac{\delta\Sigma}{\delta N^{a}_{\mu j}}
-\bar{N}^{a}_{\mu j}\frac{\delta\Sigma}{\delta\bar{N}^{a}_{\mu i}}\bigg)\nonumber\\
&=&0\,,
\end{eqnarray}
\begin{eqnarray}
\mathcal{Q}^{ab}(\Sigma)&=&\lambda^{a}_{\mu}\frac{\delta\Sigma}{\delta\lambda^{b}_{\mu}}
+\tau^{a}_{\mu}\frac{\delta\Sigma}{\delta\tau^{b}_{\mu}}
+H^{ca}\frac{\delta\Sigma}{\delta H^{cb}}
+\widetilde{H}^{ca}\frac{\delta\Sigma}{\delta \widetilde{H}^{cb}}
+T^{ca}\frac{\delta\Sigma}{\delta T^{cb}}
+\widetilde{T}^{ca}\frac{\delta\Sigma}{\delta \widetilde{T}^{cb}}\nonumber\\
&&
-X^{cb}\frac{\delta\Sigma}{\delta X^{ca}}
-\widetilde{X}^{cb}\frac{\delta\Sigma}{\delta \widetilde{X}^{ca}}
-Y^{cb}\frac{\delta\Sigma}{\delta Y^{ca}}
-\widetilde{Y}^{cb}\frac{\delta\Sigma}{\delta \widetilde{Y}^{ca}}\nonumber\\
&=&-H^{ba}-\widetilde{H}^{ba}\,,
\end{eqnarray}
\begin{eqnarray}
\mathcal{N}_{gh}(\Sigma)&=&\int d^{4}x\,\bigg(
c^{a}\frac{\delta\Sigma}{\delta c^{a}}
+\omega^{a}_{i}\frac{\delta\Sigma}{\delta \omega^{a}_{i}}
+\eta^{a}_{\mu}\frac{\delta\Sigma}{\delta \eta^{a}_{\mu}}
+u^{a}\frac{\delta\Sigma}{\delta u^{a}}
+N^{a}_{\mu i}\frac{\delta\Sigma}{\delta N^{a}_{\mu i}}
+Y^{ab}\frac{\delta\Sigma}{\delta Y^{ab}}\nonumber\\
&&
+\widetilde{Y}^{ab}\frac{\delta\Sigma}{\delta \widetilde{Y}^{ab}}
-\bar{c}^{a}\frac{\delta\Sigma}{\delta \bar{c}^{a}}
-\bar\omega^{a}_{i}\frac{\delta\Sigma}{\delta \bar\omega^{a}_{i}}
-\tau^{a}_{\mu}\frac{\delta\Sigma}{\delta \tau^{a}_{\mu}}
-\Omega^{a}_{\mu}\frac{\delta\Sigma}{\delta \Omega^{a}_{\mu}}
-\vartheta^{a}\frac{\delta\Sigma}{\delta \vartheta^{a}}
-\bar{N}^{a}_{\mu i}\frac{\delta\Sigma}{\delta \bar{N}^{a}_{\mu i}}\nonumber\\
&&
-T^{ab}\frac{\delta\Sigma}{\delta T^{ab}}
-\widetilde{T}^{ab}\frac{\delta\Sigma}{\delta \widetilde{T}^{ab}}
-2L^{a}\frac{\delta\Sigma}{\delta L^{a}}\bigg)\nonumber\\
&=&0\,.
\end{eqnarray}
}
\end{itemize}
The operator $\mathcal{N}_{gh}$ defines, in a functional way, the ghost number. Also, taking the trace in color space of $\mathcal{Q}_{ij}$, namely 
$$\mathrm{tr}[\mathcal{Q}_{ij}]\equiv \mathcal{Q}_{4(N^{2}-1)}\,,$$ defines the charge  $q_{4(N^{2}-1)}$. Analogously, taking the trace of   $\mathcal{Q}_{ab}$, {\it i.e.}  $$\mathrm{tr}[\mathcal{Q}_{ab}]\equiv \mathcal{Q}\,,$$ defines the $\mathcal{Q}$-charge. The quantum numbers of all fields and sources corresponding to these charges are displayed in Appendix~B together with the ghost number and respective dimensions.

\section{Renormalization}

Having established the Ward identities fulfilled by the action $\Sigma$, we can now study the issue of the renormalizability. We shall proceed by first analyzing the particular case in which the Gribov mass parameter $\gamma^2$ is set to zero. As we shall see, the analysis of the limit  $\gamma^2=0$ will turn out to be very helpful in the study of the general case in which $\gamma^2 \neq 0$.

\subsection{The case $\gamma^{2}=0$}

The limit $\gamma^{2}=0$ is easily achieved by setting the sources $\{M,\bar{M},N,\bar{N}\}$ to zero in the action $\Sigma$. As a consequence, the integration over the auxiliary fields $\{\varphi,\bar\varphi,\omega,\bar\omega\}$ is easily seen to give a unity. The fields $\{X,Y,\widetilde{X},\widetilde{Y}\}$ become now unnecessary. Also, the Lagrange multipliers $H$ and $\widetilde{H}$ can be integrated out. Therefore, the action $\Sigma$ in the limit $\gamma^{2}=0$ is given by:
\begin{eqnarray}
\Sigma_{\gamma^{2}=0}&\equiv&S_0\nonumber\\
&=&\int d^{4}x\,\bigg\{\frac{1}{4}\,F^{a}_{\mu\nu}F^{a}_{\mu\nu}
+\frac{\alpha}{2}\,b^{a}b^{a}+ib^{a}\,\partial_{\mu}A^{a}_{\mu}
+\bar{c}^{a}\,\partial_{\mu}D^{ab}_{\mu}(A)c^{b}
+i\rho^{a}\,\partial_{\mu}B^{a}_{\mu}
-\vartheta^{a}\,\partial_{\mu}\eta^{a}_{\mu}\nonumber\\
\phantom{\Big|}&&
+\lambda^{a}_{\mu}(B^{a}_{\mu}-(A^{a}_{\mu}-\partial_{\mu}\xi^{a}))
-\tau^{a}_{\mu}(\eta^{a}_{\mu}+\partial_{\mu}u^{a}+D^{ab}_{\mu}(A)c^{b})
-\Omega^{a}_{\mu}\,D^{ab}_{\mu}(A)c^{b}+\frac{g}{2}f^{abc}L^{a}c^{b}c^{c}\bigg\}\,.\nonumber\\
\label{gamma_zero}
\end{eqnarray}
As expected, expression  \eqref{gamma_zero} enjoys a huge set of Ward identities  which guarantee in fact its renormalizability. The BRST symmetry can be translated  in a functional form by means of the  Slavnov-Taylor identity:
\begin{equation}
\mathcal{S}(S_0)=\int d^{4}x\,\bigg(
\frac{\delta S_{0}}{\delta\Omega^{a}_{\mu}}
\frac{\delta S_{0}}{\delta A^{a}_{\mu}}
+\frac{\delta S_{0}}{\delta L^{a}}
\frac{\delta S_{0}}{\delta c^{a}}
+ib^{a}\,\frac{\delta S_{0}}{\delta\bar{c}^{a}}
+i\rho^{a}\,\frac{\delta S_{0}}{\delta\vartheta^{a}}
+\lambda^{a}_{\mu}\,\frac{\delta S_{0}}{\delta\tau^{a}_{\mu}}
+\eta^{a}_{\mu}\,\frac{\delta S_{0}}{\delta B^{a}_{\mu}}
+u^{a}\,\frac{\delta S_{0}}{\delta\xi^{a}}\bigg)=0\,.
\end{equation} 
Also, the equations of motion of $\{b,\rho,\vartheta,\lambda,\eta,u,B,\xi\}$ are linear in the fields, implying thus that the counterterm must be independent from  these fields. Moreover,   $\{\Omega,\tau,\bar{c}\}$ appear only in the combination $\hat{\Omega}=\Omega_{\mu}+\tau_{\mu}+\partial_{\mu}\bar{c}$. Therefore, following the procedure of the algebraic renormalization \cite{Piguet:1995er}, the corresponding invariant counterterm, {\it i.e.} the most general integrated local polynomial in the fields and sources with dimension four and ghost number zero compatible with all symmetries which can be freely added at each order of perturbation theory,  is given by
\begin{equation}
S^{\mathrm{CT}}_{\gamma^{2}=0}=a_{0}\,S_{\mathrm{YM}}+\mathcal{S}_{S_0}\int d^{4}x\,\left( a_{1}\,\hat\Omega^{a}_{\mu}\,A^{a}_{\mu}
+a_2\,L^{a}c^{a}\right)\,,
\label{CT_gamma_zero}
\end{equation}
where
\begin{eqnarray}
\mathcal{S}_{S_0}&=&\int d^{4}x\,\bigg(
\frac{\delta S_{0}}{\delta\Omega^{a}_{\mu}}
\frac{\delta}{\delta A^{a}_{\mu}}
+\frac{\delta S_{0}}{\delta A^{a}_{\mu}}
\frac{\delta}{\delta\Omega^{a}_{\mu}}
+\frac{\delta S_{0}}{\delta L^{a}}
\frac{\delta}{\delta c^{a}}
+\frac{\delta S_{0}}{\delta c^{a}}
\frac{\delta }{\delta L^{a}}
+ib^{a}\,\frac{\delta }{\delta\bar{c}^{a}}\nonumber\\
&&
+i\rho^{a}\,\frac{\delta }{\delta\vartheta^{a}}
+\lambda^{a}_{\mu}\,\frac{\delta }{\delta\tau^{a}_{\mu}}
+\eta^{a}_{\mu}\,\frac{\delta }{\delta B^{a}_{\mu}}
+u^{a}\,\frac{\delta }{\delta\xi^{a}}\bigg)\,,
\end{eqnarray}
is the linearized nilpotent Slavnov-Taylor operator \cite{Piguet:1995er} and $\{a_0,a_1,a_2\}$ are three independent arbitrary coefficients. \\\\ As one can easily check, the counterterm \eqref{CT_gamma_zero} can be reabsorbed into the classical action $S_0$ by a multiplicative renormalization of fields, sources and parameters, namely 
\begin{eqnarray}
S_0[\mathcal{F}_0,\mathcal{J}_0]+O(\epsilon^{2})&=& S_{0}[\mathcal{F},\mathcal{J}]+\epsilon\,S^{\mathrm{CT}}_{\gamma^{2}=0}[\mathcal{F},\mathcal{J}]\,,\nonumber\\
\mathcal{F}&=&\{A,B,\eta,\xi,u,b,c,\bar{c},\lambda,\tau,\rho,\vartheta\}\,,\nonumber\\     
\mathcal{J}&=&\{\Omega,L,g,\alpha\}\,,    \label{reabs}
\end{eqnarray}
where, the label ``0'' denotes the bare  quantities and where $\epsilon$ stands for an infinitesimal parameter expansion. Choosing the renormalization factors in the standard way, {\it i.e.}
\begin{eqnarray}
\mathcal{F}_0&=&Z^{1/2}_{\mathcal{F}}\,\mathcal{F}=\left(1+\frac{\epsilon}{2}\,z_{\mathcal{F}}\right)\mathcal{F}\,,\nonumber\\
\mathcal{J}_0&=&Z_{\mathcal{J}}\,\mathcal{J}=\left(1+\epsilon\,z_{\mathcal{J}}\right)\mathcal{J}\,, 
\end{eqnarray} 
form direct inspection of eqs.\eqref{reabs} one  finds:
\begin{eqnarray}
Z^{1/2}_{A}&=&1+\epsilon\left(\frac{a_{0}}{2}+a_1\right)\,,\nonumber\\
Z_g&=&1-\epsilon\,\frac{a_{0}}{2}\,,\nonumber\\
Z_{c}^{1/2}&=&1-\epsilon\,\frac{a_{1}+a_{2}}{2}\,,
\label{indepZs}
\end{eqnarray}
and
\begin{eqnarray}
&&Z_B=Z_{\xi}=Z_A\,,\nonumber\\
&&Z_{\lambda}=Z_{\rho}=Z_{A}^{-1}\,,\nonumber\\
&&Z_{\tau}=Z_{\vartheta}=Z_{c}\,,\nonumber\\
&&Z_{\eta}=Z_{u}=Z_{c}^{-1}\,.
\label{Zs}
\end{eqnarray}
Notice that the renormalization factors of the fields $(B, \lambda, \rho, \tau, \vartheta, \eta, u)$ are not independent, being expressed in terms of the factors $(Z_A, Z_c)$. This is not unexpected  since the action $S_0$ is nothing but the Yang-Mills action in linear covariant gauges.  These fields define in a local and off-shell way the transverse and longitudinal components of the gauge field. These extra terms are strictly necessary only in the case when $\gamma^{2}\neq 0$. However, this particular case  is useful  in order to establish a kind of boundary condition for the renormalization of the general action $\Sigma$, meaning that the renormalization factors of $\Sigma$ have to reduce to those of $S_0$ in the case in which the Gribov parameter $\gamma^2$ is set to zero. 

\subsection{General case}
Let us face now the issue of the proof of the renormalizability of the more general  action $\Sigma$. In this case, according to the algebraic renormalization set up \cite{Piguet:1995er}, for the most general invariant local counterterm we get 
\begin{equation}
\Sigma_{\mathrm{CT}}=a_{0}\,S_{\mathrm{YM}}+\mathcal{S}_{\Sigma}\Delta^{(-1)}\,,     \label{ctgz} 
\end{equation}
where $\mathcal{S}_{\Sigma}$ is the nilpotent  linearized Slavnov-Taylor operator \cite{Piguet:1995er},
\begin{eqnarray}
\mathcal{S}_{\Sigma}&=&\int d^{4}x\,\bigg(
\frac{\delta\Sigma}{\delta\Omega^{a}_{\mu}}\frac{\delta}{\delta A^{a}_{\mu}}
+\frac{\delta\Sigma}{\delta A^{a}_{\mu}}\frac{\delta}{\delta \Omega^{a}_{\mu}}
+\frac{\delta\Sigma}{\delta L^{a}}\frac{\delta}{\delta c^{a}}
+\frac{\delta\Sigma}{\delta c^{a}}\frac{\delta}{\delta L^{a}}
+i{b}^{a}\frac{\delta}{\delta\bar{c}^{a}}
+\eta^{a}_{\mu}\frac{\delta}{\delta B^{a}_{\mu}}\nonumber\\
&+&i\rho^{a}\frac{\delta}{\delta \vartheta^{a}}
+u^{a}\frac{\delta}{\delta \xi^{a}}
+\omega^{a}_{i}\frac{\delta}{\delta \varphi^{a}_{i}}
+\bar\varphi^{a}_{i}\frac{\delta}{\delta\bar\omega^{a}_{i}}
+N^{a}_{\mu i}\frac{\delta}{\delta M^{a}_{\mu i}}
+\bar{M}^{a}_{\mu i}\frac{\delta}{\delta\bar{N}^{a}_{\mu i}}
+\lambda^{a}_{\mu}\frac{\delta}{\delta\tau^{a}_{\mu}}\nonumber\\
&+&Y^{ab}\frac{\delta}{\delta X^{ab}}
+\widetilde{Y}^{ab}\frac{\delta}{\delta \widetilde{X}^{ab}}
+H^{ab}\frac{\delta}{\delta T^{ab}}
+\widetilde{H}^{ab}\frac{\delta}{\delta \widetilde{T}^{ab}}\,\bigg)\,,
\end{eqnarray}
and $\Delta^{(-1)}$ is an integrated local polynomial in the fields with dimension $4$, ghost number $(-1)$ and with vanishing $\mathcal{Q}$ and $q_{4(N^{2}-1)}$ charges. Taking into account the full set of Ward identities derived in the last section, it turns out that, after a lengthy algebraic analysis, the term  $\Delta^{(-1)}$ can be written as 
\begin{eqnarray}
\Delta^{(-1)} &=& \int d^{4}x \,\bigg[a_{1}\left(\Omega_{\mu}^{a} +\partial_{\mu}^{a}\bar c^{a} + \widetilde{X}^{ab}\tau_{\mu}^{b}\right)A_{\mu}^{a}  + a_{2}c^{a}L^{a} \nonumber \\
&+& a_{3}\left(M_{\mu i}^{a}\bar{N}_{\mu i}^{a} + M_{\mu i}^{a}\,\partial_{\mu}\bar\omega_{i}^{a} +\bar{N}_{\mu i}^{a}\, \partial_{\mu}\varphi ^{a}_{i} -\varphi ^{a}_{i}\partial^{2}\bar\omega ^{a}_{i}\right)\,\bigg]\,.
\label{Delta}
\end{eqnarray}
Notice that terms like $(\Omega_{\mu}^{a} +\partial_{\mu}^{a}\bar c^{a} + \widetilde{X}^{ab}\tau_{\mu}^{b})B_{\mu}^{a}$ are forbidden due to the Ward identity  \eqref{W_symmetry}. It remains now to check out if the countertem above can be reabsorbed into the classical starting action $\Sigma$ though a  multiplicative renormalization of the fields, sources and parameters, namely 
\begin{equation}
\Sigma[\mathcal{F}_0,\mathcal{J}_0]+O(\epsilon^{2})=\Sigma[\mathcal{F},\mathcal{J}]+\epsilon\,\Sigma_{\mathrm{CT}}\,.
\end{equation}
where, as before, $\epsilon$ denotes the expansion parameter,  $\mathcal{F}$ stands for the fields and $\mathcal{J}$ for  the external sources and parameters. Setting 
\begin{equation}
\mathcal{F}_0=Z^{1/2}_{\mathcal{F}}\,\mathcal{F}=\left(1+\frac{\epsilon}{2}\,z_{\mathcal{F}}\right)\,\mathcal{F}\,,\qquad
\mathcal{J}_0=Z_{\mathcal{J}}\,\mathcal{J}=\left(1+\epsilon\,z_{\mathcal{J}}\right)\,\mathcal{J}\,,
\end{equation}
with $(z_{\mathcal{F}},z_{\mathcal{J}})$ being linear combinations of the dimensionless coefficients $(a_{0},a_{1},a_{2},a_{3})$, it turns out that 
\begin{eqnarray}
Z^{1/2}_{A}&=&1+\epsilon\left(\frac{a_{0}}{2}+a_1\right)\,,\nonumber\\
Z^{1/2}_{B}&=&1+\epsilon\left(\frac{a_{0}}{2}+a_3\right)\,,\nonumber\\
Z_g&=&1-\epsilon\,\frac{a_{0}}{2}\,,\nonumber\\
Z_{c}^{1/2}&=&1-\epsilon\,\frac{a_{1}+a_{2}}{2}\,. 
\label{indepZs2}
\end{eqnarray}
In order to determine the remaining renormalization factors, it is helpful to notice that the coefficients  $(a_{0},a_{1},a_{2},a_{3})$ are independent from the external sources $(M,N, {\bar M}, {\bar N})$. Moreover, when these external sources are set to zero, {\it i.e.} $(M,N, {\bar M}, {\bar N})=0$ they have to reduce to the renormalization factors encountered before in the analysis of the action $S_0$, since setting $(M,N, {\bar M}, {\bar N})=0$ is equivalent to take the limit $\gamma^2=0$, eq.\eqref{plm}. This immediately gives  $a_3=a_1$, implying that $Z_{B}=Z_{A}$. Therefore, we obtain
\begin{eqnarray}
&&Z_B=Z_{\xi}=Z_{\alpha}=Z_L=Z_A\,,\nonumber\\
&&Z_{\lambda}=Z_{\rho}=Z_{b}=Z_{A}^{-1}\,,\nonumber\\
&&Z_{\bar{c}}=Z_{\tau}=Z_{\vartheta}=Z_{K}=Z_{c}\,,\nonumber\\
&&Z_{\eta}=Z_{v}=Z_{u}=Z_{c}^{-1}\,,\nonumber\\
&&Z^{1/2}_{\varphi}=Z^{1/2}_{\bar{\varphi}}=Z_{M}=Z_{\bar{M}}=Z_{g}^{-1/2}Z_{A}^{-1/4}\,,\nonumber\\
&&Z^{1/2}_{\bar\omega}=Z_{\bar{N}}=Z_{c}^{1/2}Z_{g}^{-1/2}Z_{A}^{1/4}\,,\nonumber\\
&&Z^{1/2}_{\omega}=Z_{N}=Z_{c}^{-1/2}Z_{g}^{-1/2}Z_{A}^{-3/4}\,,\nonumber\\
&&Z_X=Z_{\widetilde{X}}=Z_{H}=Z_{\widetilde{H}}=1\,,\nonumber\\
&&Z_{Y}^{1/2}=Z_{\widetilde{Y}}^{1/2}=Z^{-1/2}_{T}=Z^{-1/2}_{\widetilde{T}}=Z_{c}^{-1/2}Z^{-1/2}_{A}\,,\nonumber\\
&&Z_{\Omega}=Z_{c}^{1/2}\,,\qquad Z_{L}=Z_{A}^{1/2}\,,
\label{depZs}
\end{eqnarray}
while, as expected, it also follows that  $\kappa=-1$.\\\\This ends the proof of the all order multiplicative renormalization of the Gribov-Zwanziger action in the linear covariant gauges  in the approximation $\mathbf{A}_{\mu}=A^{T}_{\mu}$.

\section{The refinement of the Gribov-Zwanziger action in the linear covariant gauges: introduction of dimension two condensates}

As already mentioned in the introduction, the Gribov-Zwanziger set up naturally generalizes to the so-called refined Gribov-Zwanziger action   \cite{Capri:2015pja,Capri:2015nzw,Dudal:2008sp,Capri:2015pfa,Guimaraes:2015bra} in which further non-perturbative effects, encoded in the formation of dimension two condensates, are taken into account.  Therefore, according to \cite{Capri:2015pja,Capri:2015nzw,Dudal:2008sp,Capri:2015pfa,Guimaraes:2015bra}, in order to extend the previous results to the refined version  we have to properly introduce the dimension two operators:
\begin{equation}
\mathcal{O}_{1}=\left(\bar\varphi^{a}_{i}\varphi^{a}_{i}-\bar\omega^{a}_{i}\omega^{a}_{i}\right)\,,\qquad \mathcal{O}_{2}=A^{T}_{\mu}A^{T}_{\mu}\,.  \label{o1o2}
\end{equation}
The introduction of operator $\mathcal{O}_{1}$ is immediate because it is BRST invariant. The inclusion of the operator $\mathcal{O}_{2}$ is more subtle. First of all, it needs to be introduced with the help of the off-shell and local version of $A^{T}_{\mu}$, {\it i.e.} through  the field $B^{a}_{\mu}$. In other words, we have to introduce the operator $\mathcal{O}_{2}$  written as $B^{a}_{\mu}B^{a}_{\mu}$, where the field $B^a_\mu$, eqs.\eqref{lambda_constraint},\eqref{B}, is subject to the constraints $\partial_{\mu}B^{a}_{\mu}=0$ and $A^{a}_{\mu}=B^{a}_{\mu}+\partial_{\mu}\xi^{a}$, implemented by means of the  Lagrange multipliers $\rho^{a}$ and $\lambda^{a}_{\mu}$, respectively. The second  non-trivial point we have to deal is to find a way to introduce the operator $B^{a}_{\mu}B^{a}_{\mu}$, while  forbidding the potential mixing with the operators  $A^{a}_{\mu}A^{a}_{\mu}$ and $A^{a}_{\mu}B^{a}_{\mu}$. 

A consistent introduction of the operator $B^{a}_{\mu}B^{a}_{\mu}$ is achieved by extending the Ward identities  \eqref{W_symmetry} and \eqref{Wbar_symmetry} by means of a BRST doublet  of external sources $(U^{abc}, V^{abc})$ carrying  color indices, namely 
\begin{equation}
sU^{abc}=V^{abc}\,,\qquad sV^{abc}=0\,.
\end{equation}
The physical values of these sources are
\begin{equation}
V^{abc}\,\big|_{\mathrm{phys}}=\frac{m^{2}}{2N}\,f^{abc}\,,\qquad U^{abc}\,\big|_{\mathrm{phys}}=0   \;. 
\end{equation}
Thus, we add to the action $\Sigma$, eq.\eqref{action}, the term:
\begin{eqnarray}
S_{\mathcal{O}_{2}}&=&s\int d^{4}x\,U^{abc}\,f^{abd}B^{d}_{\mu}B^{c}_{\mu}\nonumber\\
&=&\int d^{4}x\, \left(V^{abc}\,f^{abd}B^{d}_{\mu}B^{c}_{\mu}-U^{abc}\,f^{abd}\eta^{d}_{\mu}B^{c}_{\mu}
-U^{abc}\,f^{abd}B^{d}_{\mu}\eta^{c}_{\mu}\right)\,.
\end{eqnarray}     
Notice that, in the physical limit, we have
\begin{equation}
S_{\mathcal{O}_{2}}\,\big|_{\mathrm{phys}}=\int d^{4}x\,\frac{m^{2}}{2N}\,\underbrace{f^{abc}f^{abd}}_{N\delta^{cd}}B^{d}_{\mu}B^{c}_{\mu}=\int d^{4}x\,\frac{m^{2}}{2}\,B^{a}_{\mu}B^{a}_{\mu}\,,
\end{equation}
which nicely shows that the operator $B^{a}_{\mu}B^{a}_{\mu}$ is correctly obtained. Moreover, the Ward operators $\mathcal{W}^{a}$ and $\overline{\mathcal{W}}^{a}$ generalize to
\begin{eqnarray}
\mathcal{W}^{a}\,\to\, \mathcal{W}^{a}_{\mathrm{new}}&\!\!=\!\!&\mathcal{W}^{a}+f^{abc}\int d^{4}x\,\bigg(V^{bmn}\frac{\delta}{\delta V^{cmn}}
+V^{mbn}\frac{\delta}{\delta V^{mcn}}
+V^{mnb}\frac{\delta}{\delta V^{mnc}}\nonumber\\
&+&U^{bmn}\frac{\delta}{\delta U^{cmn}}
+U^{mbn}\frac{\delta}{\delta U^{mcn}}
+U^{mnb}\frac{\delta}{\delta U^{mnc}}\bigg)\,,\nonumber\\
\overline{\mathcal{W}}^{a}\,\to\, \overline{\mathcal{W}}^{a}_{\mathrm{new}}&\!\!=\!\!&\overline{\mathcal{W}}^{a}+f^{abc}\int d^{4}x\,\bigg(
U^{bmn}\frac{\delta}{\delta V^{cmn}}
+U^{mbn}\frac{\delta}{\delta V^{mcn}}
+U^{mnb}\frac{\delta}{\delta V^{mnc}}\bigg)\,.
\label{3nd_news}
\end{eqnarray} 
It is easy to check that 
\begin{equation}
\mathcal{W}^{a}_{\mathrm{new}}(S_{\mathcal{O}_{2}})=0\,,\qquad
\overline{\mathcal{W}}^{a}_{\mathrm{new}}(S_{\mathcal{O}_{2}})=0\,,    \label{ww}
\end{equation}
showing that the term $S_{\mathcal{O}_{2}}$ is in fact left invariant. As a consequence of eqs.\eqref{ww}, unwanted mixing terms like 
 $V^{abc}f^{abd}A^{c}_{\mu}A^{d}_{\mu}$ and $V^{abc}f^{abd}A^{c}_{\mu}B^{d}_{\mu}$ turn out to be forbidden. The same happens for terms of the kind $V^{abc}f^{abd}X^{cm}X^{dn}A^{m}_{\mu}A^{n}_{\mu}$, which are ruled out by the quantum numbers of the fields and sources, see Appendix B.

We can turn to the operator $\mathcal{O}_{1}$. It can be introduced trough  a BRST invariant source, $\Lambda$, according to 
\begin{equation}
S_{\mathcal{O}_{1}}=s\int d^{4}x\,\Lambda\,\bar{\omega}^{a}_{i}\varphi^{a}_{i}
=\int d^{4}x\,\Lambda\,(\bar\varphi^{a}_{i}\varphi^{a}_{i}-\bar\omega^{a}_{i}\omega^{a}_{i})\,.
\end{equation}
At the end, the source $\Lambda$ is set to its physical value:
\begin{equation}
\Lambda\,\big|_{\mathrm{phys}}=\mu^{2}\,,
\end{equation} 
which gives the inclusion of the desired operator $\mathcal{O}_{1}$. In turn, terms like $\Lambda\,A^{a}_{\mu}A^{a}_{\mu}$ and $\Lambda\,B^{a}_{\mu}B^{a}_{\mu}$ are forbidden by the BRST symmetry, as it follows by noticing that they are not BRST invariant. 

Thus, the new extended action containing the relevant dimension two operators is:
\begin{eqnarray}
\Sigma_{\mathrm{new}}&=&\Sigma+s\int d^4x\,\bigg(\Lambda\,\bar\omega^{a}_{i}\varphi^{a}_{i}
+U^{abc}\,f^{abd} B^{c}_{\mu}B^{d}_{\mu}
+\zeta^{abcdef}_{1}\,U^{abc}V^{def}+\zeta_{2}\,f^{abc}U^{abc}\Lambda\,\bigg)\nonumber\\
&&
+\zeta_{3}\,\int d^{4}x\,\Lambda^{2}\nonumber\\
&=&\Sigma+\int d^{4}x\,\bigg[\Lambda(\bar\varphi^{a}_{i}\varphi^{a}_{i}-\bar\omega^{a}_{i}\omega^{a}_{i})
+V^{abc}\,f^{abd}B^{c}_{\mu}B^{d}_{\mu}\nonumber\\
&&-U^{abc}f^{abd}(\eta^{c}_{\mu}B^{d}_{\mu}+\eta^{d}_{\mu}B^{c}_{\mu})
+\zeta_{1}^{abcdef}\,V^{abc}V^{def}
+\zeta_{2}\,f^{abc}V^{abc}\Lambda
+\zeta_{3}\,\Lambda^{2}\,\bigg]\,.    \label{ref}
\end{eqnarray} 
The terms in the external sources $(\zeta_{1}^{abcdef}\,V^{abc}V^{def}
+\zeta_{2}\,f^{abc}V^{abc}\Lambda
+\zeta_{3}\,\Lambda^{2}$  are vacuum terms, allowed by power counting. Also, the dimensionless coefficient $\zeta_{1}^{abcde}$  fulfils the conditions:
\begin{eqnarray}
&&\zeta_{1}^{abcdef}=\zeta_{1}^{defabc}\,,\nonumber\\
&&\zeta_{1}^{cmndef}f^{abc}
+\zeta_{1}^{bcndef}f^{amc}
+\zeta_{1}^{bmcdef}f^{anc}=0\,.
\end{eqnarray}
The inclusion  of the dimension two operators $(\mathcal{O}_{1},\mathcal{O}_{2})$ does not invalidate the Ward identities corresponding to the equations of motions of the localizing Zwanziger fields $\{\varphi,\bar\varphi,\omega,\bar\omega\}$, which takes now the form  
\begin{eqnarray}
\frac{\delta\Sigma_{\mathrm{new}}}{\delta\bar\varphi^{a}_{i}}+\partial_{\mu}\frac{\delta\Sigma_{\mathrm{new}}}{\delta\bar{M}^{a}_{\mu i}}&\!\!\!=\!\!\!&-gf^{abc}M^{b}_{\mu i}B^{c}_{\mu}+\Lambda\,\varphi^{a}_{i}\,,\nonumber\\
\frac{\delta\Sigma_{\mathrm{new}}}{\delta\bar\omega^{a}_{i}}+\partial_{\mu}\frac{\delta\Sigma_{\mathrm{new}}}{\delta\bar{N}^{a}_{\mu i}}&=&gf^{abc}N^{b}_{\mu i}B^{c}_{\mu}+gf^{abc}M^{b}_{\mu i}\eta^{c}_{\mu}-\Lambda\,\omega^{a}_{i}\,,\nonumber\\
\frac{\delta\Sigma_{\mathrm{new}}}{\delta\omega^{a}_{i}}+\partial_{\mu}\frac{\delta\Sigma_{\mathrm{new}}}{\delta{N}^{a}_{\mu i}}+igf^{abc}\bar\omega^{b}_{i}\frac{\delta\Sigma_{\mathrm{new}}}{\delta\rho^{c}}
&\!\!\!=\!\!\!&-gf^{abc}\bar{N}^{b}_{\mu i}B^{c}_{\mu}+\Lambda\,\bar\omega^{a}_{i}\,,\nonumber\\
\!\!\!\!\!\!\frac{\delta\Sigma_{\mathrm{new}}}{\delta\varphi^{a}_{i}}+\partial_{\mu}\frac{\delta\Sigma_{\mathrm{new}}}{\delta{M}^{a}_{\mu i}}+igf^{abc}\bar\varphi^{b}_{i}\frac{\delta\Sigma_{\mathrm{new}}}{\delta\rho^{c}}
-gf^{abc}\bar\omega^{b}_{i}\frac{\delta\Sigma_{\mathrm{new}}}{\delta\vartheta^{c}}
&\!\!\!=\!\!\!&-gf^{abc}\bar{M}^{b}_{\mu i}B^{c}_{\mu}
+gf^{abc}\bar{N}^{b}_{\mu i}\eta^{c}_{\mu}+\Lambda\,\bar\varphi^{a}_{i}\,,\nonumber\\
\label{eqs_zwanziger_fields_new}
\end{eqnarray} 
from which one sees that the inclusion of $(\mathcal{O}_{1},\mathcal{O}_{2})$ yields only additional harmless linear breaking terms in the left hand-sides of \eqref{eqs_zwanziger_fields_new}. 

Repeating now the previous algebraic analysis, it follows that the most general local invariant counterterm in the presence of the operators $(\mathcal{O}_{1},\mathcal{O}_{2})$ is given by 
\begin{eqnarray}
\Sigma^{\mathrm{new}}_{\mathrm{CT}}&=&\Sigma_{\mathrm{CT}}+
\int d^{4}x\,\bigg[
\lambda_{1}^{abcde}\,V^{abc}B^{d}_{\mu}B^{e}_{\mu}-\lambda_{1}^{abcde}\,U^{abc}(\eta^{d}_{\mu}B^{e}_{\mu}+\eta^{e}_{\mu}B^{d}_{\mu})
\nonumber\\
&+&\lambda_{2}^{abcdef}\,V^{abc}V^{def}
+\lambda_{3}\,\zeta_{2}f^{abc}\,V^{abc}\Lambda
+\lambda_{4}\,\zeta_{3}\,\Lambda^{2}\,\bigg]\,, 
\end{eqnarray}
where $(\lambda_{1}^{abcde}, \lambda_{2}^{abcdef}, \lambda_{3},\lambda_{4})$ are free coefficients and where  $\Sigma_{\mathrm{CT}}$ is given by eq.\eqref{ctgz}. Similarly to the quantity $\zeta_{1}^{abcdef}$, the coefficients $\lambda_{1}^{abcde}$ and $\lambda_{2}^{abcdef}$ obey the identities:
\begin{eqnarray}
&&\lambda_{1}^{cmnde}f^{abc}+\lambda_{1}^{bcnde}f^{amc}
+\lambda_{1}^{bmcde}f^{anc}+\lambda_{1}^{bmnce}f^{adc}
+\lambda_{1}^{bmndc}f^{aec}=0\,,\nonumber\\
&&\lambda_{2}^{abcdef}=\lambda_{2}^{defabc}\,,\nonumber\\
&&\lambda_{2}^{cmndef}f^{abc}
+\lambda_{2}^{bcndef}f^{amc}
+\lambda_{2}^{bmcdef}f^{anc}=0\,.
\end{eqnarray}
In particular, we point out that terms like $\Lambda\,(\bar\varphi^{a}_{i}\varphi^{a}_{i}-\bar\omega^{a}_{i}\omega^{a}_{i})$ are forbidden by the linearly broken Ward identities \eqref{eqs_zwanziger_fields_new}.\\\\The new sources and parameters are easily seen to renormalize as
\begin{eqnarray}
V_{0}^{abc}&=&V^{abc}+\epsilon\,z_{V}^{abcdef}\,V^{def}\,,\nonumber\\
U_{0}^{abc}&=&U^{abc}+\epsilon\,z_{U}^{abcdef}\,U^{def}\,,\nonumber\\
\Lambda_{0}&=&Z_{\Lambda}\,\Lambda\,,\nonumber\\
\zeta_{1,0}^{abcdef}&=&\zeta_{1}^{abcdef}+\epsilon\,z^{abcdef}\,,\nonumber\\
\zeta_{2,0}&=&Z_{\zeta_{2}}\,\zeta_{2}\,,\nonumber\\
\zeta_{3,0}&=&Z_{\zeta_{3}}\,\zeta_{3}\,.
\end{eqnarray}
Therefore, the whole counterterm $\Sigma^{\mathrm{new}}_{\mathrm{CT}}$ is reabsorbed into the starting action $\Sigma_{\mathrm{new}}$  through the following  redefinitions
\begin{eqnarray}
f^{abh}z_{V}^{abcdef}&=&\lambda_{1}^{defch}-(a_0+2a_1)f^{deh}\delta^{cf}\,,\nonumber\\
f^{abh}z_{U}^{abcdef}&=&\lambda_{1}^{defch}-\frac{a_0+3a_1+a_2}{2}f^{deh}\delta^{cf}\,,\nonumber\\
Z_{\Lambda}&=&Z_{\varphi}^{-1}=Z_{g}Z_{A}^{1/2}\,,\nonumber\\
z^{abcdef}&=&\lambda_{2}^{abcdef}-2z_{V}^{pqrabc}\,\zeta_{1}^{pqrdef}\,,\nonumber\\
Z_{\zeta_{2}}&=&1+\epsilon\,(\lambda_{3}-\lambda_{1}+a_{0}+a_{1})\,,\nonumber\\
Z_{\zeta_{3}}&=&1+\epsilon\,(\lambda_{4}-a_{1})\,,
\end{eqnarray}
where
\begin{equation}
\lambda_{1}\equiv\frac{f^{abc}\lambda_{1}^{abcdd}}{N(N^{2}-1)}\,.
\end{equation}
Taking the physical values of the sources, one obtains that the parameters $m^{2}$ and $\mu^{2}$ defining the dimension two operators $(\mathcal{O}_{1},\mathcal{O}_{2})$ renormalize as
\begin{equation}
Z_{\mu^{2}}=Z_{g}Z_{A}^{1/2}\,,\qquad
Z_{m^{2}}=Z_{A}^{-1}(1+\epsilon\,\lambda_{1})\,.
\end{equation}
In particular, from the presence of the free coefficient $\lambda_{1}$ in the expression for $Z_{m^{2}}$, it follows that, in the case of the linear covariant gauges, the renormalization factor of the operator $\langle A^{T}_{\mu}A^{T}_{\mu}\rangle$ is an independent parameter of the theory. This result is in contrast with the case of the Landau gauge, in which the renormalization  factors of the parameters  $m^{2}$ and $\mu^{2}$ can be expressed solely in terms of $Z_g$ and $Z_A$. 

In summary,  the introduction of the dimension two operators  $(\mathcal{O}_{1},\mathcal{O}_{2})$ does not spoil the renormalizability of the model, providing thus a local and renormalizable framework to handle the refined Gribov-Zwanziger action $\Sigma_{\mathrm{new}}$  in the approximation $\mathbf{A}_{\mu}\approx A^{T}_{\mu}$.

\section{Conclusion}

In this work we have studied the Gribov-Zwanziger model in the linear covariant gauges in the approximation $\mathbf{A}_{\mu}=A^{T}_{\mu}$, amounting to make use of the horizon function expressed in terms of the transverse component of the gauge field, eq.\eqref{horizon_LCG}. The resulting action $\Sigma$, eq.\eqref{action}, has been cast in local form and has been proven to be multiplicative renormalizable to all orders of perturbation theory within the algebraic renormalization set up. 

Subsequently, the refined version of the model, given by the extended action $\Sigma_{\mathrm{new}}$ of eq.\eqref{ref}, and corresponding to the introduction of the dimensions two operators $(\mathcal{O}_{1},\mathcal{O}_{2})$ of eq.\eqref{o1o2}, has been constructed and shown to be also renormalizable. 

Both results are nontrivial and can be taken as strong indication of the possible renormalizability of the Gribov-Zwanziger framework when the full non-local  gauge invariant field $\mathbf{A}_{\mu}$ is employed  \cite{Capri:2015pja}, a task which is already under investigation \cite{progress}.

\section*{Acknowledgments}

The Conselho Nacional de Desenvolvimento Cient\'{\i}fico e
Tecnol\'{o}gico (CNPq-Brazil), the Faperj, Funda{\c{c}}{\~{a}}o de
Amparo {\`{a}} Pesquisa do Estado do Rio de Janeiro, the SR2-UERJ,  the
Coordena{\c{c}}{\~{a}}o de Aperfei{\c{c}}oamento de Pessoal de
N{\'{\i}}vel Superior (CAPES)  are gratefully acknowledged.

\appendix

\section{Tree-level propagators}
For the benefit of the reader, let us give here the propagators of the elementary fields. The quadratic part of the action $\Sigma$, eq.~\eqref{action}, taking the physical values of the sources $\{M,\bar{M},N,\bar{N},\Omega,L\}$ and integrating out the Lagrange multipliers $\{H,\widetilde{H}\}$, is given by:
\begin{eqnarray}
\Sigma_{\mathrm{quad}} &=& \int d^4x\bigg[ \frac{1}{2}\,(\partial_{\mu}A^{a}_{\nu}-\partial_{\nu}A^{a}_{\nu})\partial_{\mu}A^{a}_{\nu}+ib^{a}\,\partial_{\mu}A^{a}_{\mu}+\frac{\alpha}{2}\,b^ab^{a} 
+\bar{\varphi}^{ab}_{\mu}\partial^2\varphi^{ab}_{\mu}
+g\gamma^2f^{abc}B_{\mu}^a\,(\varphi^{bc}_{\mu}+\bar{\varphi}^{bc}_{\mu})
\nonumber \\ 
&&+\lambda^a_{\mu}\left(B^a_{\mu}-A^a_{\mu}+\partial_{\mu}\xi^a\right)
+i\rho^a\,\partial_{\mu}B_{\mu}^a\,+\bar{c}^{a}\,\partial^{2}c^{a}
-\bar{\omega}^{ab}_{\mu}\partial^2\omega^{ab}_{\mu}
-\vartheta^{a}\,\partial_{\mu}\eta^{a}_{\mu}
-\tau^{a}_{\mu}\eta^{a}_{\mu}\nonumber\\
&&-\tau^{a}_{\mu}\,\partial_{\mu}c^{a}
-\tau^{a}\,\partial_{\mu}u^{a}\bigg]
\,. 
\label{Squad}
\end{eqnarray}
Notice that the anticommuting  sector $\{c,\bar{c},\omega,\bar\omega, \vartheta,\eta,\tau,u\}$ is completely decoupled from the rest of the theory. As a consequence, the quadratic part action can be written as
\begin{equation}
\Sigma_{\mathrm{quad}}=\frac{1}{2}\int d^4x\,\Phi^{T}\,\mathbf{M}\,\Phi\,,
\end{equation}
where
\begin{equation}
\Phi^{T}=\left(\begin{array}{cccccccc}A_{\alpha}^{a} & b^{b} & B_{\beta}^c & V_{\gamma}^{de} & U_{\eta}^{fg} & \lambda_{\lambda}^{h} & \xi^{i} & \rho^j \end{array}\right)\,,\qquad
\Phi=\left(\begin{array}{c}A_{\mu}^{l}\\b^{m}\\B_{\rho}^n\\V_{\sigma}^{op}\\U_{\tau}^{qr}\\\lambda_{\omega}^{s}\\
\xi^t \\ \rho^v \end{array}\right)\,,
\end{equation}
and 
\begin{equation}
{\scriptsize
\arraycolsep=4pt 
\medmuskip = 4mu 
\mathbf{M}=\left(\begin{array}{c|c|c|c|c|c|c|c}(-\delta_{\alpha\mu}\partial^{2}+\partial_{\alpha}\partial_{\mu})\delta^{al} & -i\delta^{am}\partial_{\alpha} & 0 & 0 & 0 & -\delta^{as}\delta_{\alpha\omega}\partial^2 & 0 & 0
\\
i\delta^{bl}\partial_{\mu} & \alpha\delta^{bm} & 0 & 0 & 0 & 0 & 0 & 0
\\
0 & 0 & 0& g\gamma^{2}f^{cop}\delta_{\beta\sigma} & 0 & \delta^{cs}\delta_{\beta\omega}\partial^2 
 & 0 & -i\delta^{cv}\partial_{\beta}
\\
0 & 0 & g\gamma^{2}f^{nde}\delta_{\gamma\rho} & \frac{1}{2}\delta^{do}\delta^{ep}\delta_{\gamma\sigma}\partial^2 & 0 & 0 & 0 & 0
\\
0 & 0 & 0 & 0 & \frac{1}{2}\delta^{fq}\delta^{gr}\delta_{\eta\tau}\partial^{2} & 0 & 0 & 0
\\
-\delta^{hl}\delta_{\lambda\mu}\partial^2 & 0 & \delta^{hn}\delta_{\lambda\rho}\partial^2 
& 0 & 0 & 0 &\delta^{ht}\partial^{2}\partial_{\lambda} & 0
\\
0 & 0 & 0 & 0 & 0 & -\delta^{is}\partial^2\partial_{\omega} & 0 & 0
\\
0 & 0 & i\delta^{jn}\partial_{\rho} & 0 & 0 & 0 & 0 & 0
\end{array}\right)}
\end{equation}
with
\begin{equation}
\bar{\varphi}_{\mu}^{ab}=\frac{V_{\mu}^{ab}-iU_{\mu}^{ab}}{2}\,,\qquad
\varphi_{\mu}^{ab}=\frac{V_{\mu}^{ab}+iU_{\mu}^{ab}}{2}\,.
\end{equation}
The propagators are obtained by evaluating the inverse of the matrix $\mathbf{M}$. In momentum space, they are given by:
\begin{eqnarray}
\langle A_{\mu}^{a}\left(p\right)A_{\nu}^{b}\left(-p\right)\rangle&=&\left(\,\frac{p^{2}}{p^{4}+2Ng^{2}\gamma^{4}}{P}_{\mu\nu}+\frac{\alpha}{p^{2}}\,\frac{p_{\mu}p_{\nu}}{p^{2}}\,\right)\delta^{ab}\,,
\\
\langle\varphi_{\mu}^{ab}\left(p\right)\varphi_{\nu}^{cd}\left(-p\right)\rangle= \langle\bar{\varphi}_{\mu}^{ab}\left(p\right)\bar{\varphi}_{\nu}^{cd}\left(-p\right)\rangle & = & \frac{g^{2}\gamma^{4}f^{nab}f^{ncd}}{p^{2}\left(p^{4}+2Ng^{2}\gamma^{4}\right)}{P}_{\mu\nu}\,,
\\
\langle\varphi_{\mu}^{ab}\left(p\right)\bar{\varphi}_{\nu}^{cd}\left(-p\right)\rangle &=&\frac{g^{2}\gamma^{4}f^{nab}f^{ncd}}{p^{2}\left(p^{4}+2Ng^{2}\gamma^{4}\right)}{P}_{\mu\nu}-\frac{1}{p^{2}}\delta^{ac}\delta^{bd}\delta_{\mu\nu}\,,
\\
\langle A_{\mu}^{a}\left(p\right)\varphi_{\nu}^{bc}\left(-p\right)\rangle=
\langle A_{\mu}^{a}\left(p\right)\bar{\varphi}_{\nu}^{bc}\left(-p\right)\rangle&=&
\frac{g\gamma^{2}f^{abc}}{p^{4}+2Ng^{2}\gamma^{4}}{P}_{\mu\nu}\,,\\\cr
\langle A_{\mu}^{a}\left(p\right)b^{b}\left(-p\right)\rangle&=&\frac{1}{p^{2}}\delta^{ab}p_{\mu}\,,\\\cr
\langle b^{a}\left(p\right)\xi^{b}\left(-p\right)\rangle&=&\frac{i}{p^2}\delta^{ab}\,,\\\cr
\langle B_{\mu}^{a}(p)B_{\nu}^{b}(-p)\rangle=\langle B_{\mu}^{a}(p)A_{\nu}^{b}(-p)\rangle&=&
\frac{p^2\delta^{ab}}{p^{4}+2Ng^{2}\gamma^{4}}{P}_{\mu\nu}\,,\\\cr
\langle B_{\mu}^{a}(p)\varphi_{\nu}^{bc}(-p)\rangle =
\langle B_{\mu}^{a}(p)\bar{\varphi}_{\nu}^{bc}(-p)\rangle&=&
\frac{g\gamma^{2}f^{abc}}{p^{4}+2Ng^{2}\gamma^{4}}{P}_{\mu\nu}\,,\\\cr
\langle B_{\mu}^{a}(p)\lambda_{\nu}^{b}(-p)\rangle&=&
-\frac{p^4\delta^{ab}}{p^{4}+2Ng^{2}\gamma^{4}}{P}_{\mu\nu}\,,
\\\cr
\langle \lambda_{\mu}^{a}(p)\lambda_{\nu}^{b}(-p)\rangle & = &\frac{2Ng^2\gamma^{4}\,p^{2}\delta^{ab}}{p^{4}+2Ng^{2}\gamma^{4}}{P}_{\mu\nu}\,,
\\\cr
\langle \lambda_{\mu}^{a}(p)\varphi_{\nu}^{bc}(-p)\rangle
=\langle \lambda_{\mu}^{a}(p)\bar{\varphi}_{\nu}^{bc}(-p)\rangle & = &
-\frac{Ng\gamma^{2}\,p^{2}f^{abc}}{p^{4}+2Ng^{2}\gamma^{4}}{P}_{\mu\nu}\,,
\\\cr
\langle \lambda_{\mu}^{a}(p)A^{b}_{\nu}(-p)\rangle & = &
-\frac{2Ng^2\gamma^4}{p^4+2Ng^2\gamma^4}\delta^{ab}{P}_{\mu\nu}\,,
\\\cr
\langle\xi^{a}(p)\xi^{b}(-p)\rangle&=&\frac{\alpha}{p^4}\delta^{ab}\,,\\\cr
\langle\xi^{a}(p)A^{b}_{\mu}(-p)\rangle&=&\frac{i\alpha}{p^4}\delta^{ab}p_{\mu}\,,\\\cr
\langle\rho^{a}(p)\rho^{b}(-p)\rangle&=&\frac{2Ng^2\gamma^4}{p^4}\delta^{ab}\,,\\\cr
\langle\bar{\varphi}^{ab}_{\mu}(p)\rho^{c}(-p)\rangle=
\langle\varphi^{ab}_{\mu}(p)\rho^{c}(-p)\rangle&=&\frac{g\gamma^2}{p^4}f^{abc}p_{\mu}\,,\\\cr
\langle\xi^{a}(p)\rho^{b}(-p)\rangle&=&-\frac{i}{p^2}\delta^{ab}\,,\\\cr
\langle\xi^{a}(p)b^{b}(-p)\rangle&=&\frac{i}{p^2}\delta^{ab}\,,\\\cr
\langle\xi^{a}(p)\lambda_{\mu}^{b}(-p)\rangle&=&-\frac{i}{p^4}\delta^{ab}p_{\mu}\,,\\\cr
\langle B_{\mu}^{a}(p)\rho^{b}(-p)\rangle&=&\frac{1}{p^2}\delta^{ab}p_{\mu}\,,
\end{eqnarray}
with $P_{\mu\nu}=\delta_{\mu\nu}-\frac{p_{\mu}p_{\nu}}{p^{2}}$ being the tranverse projector.\\\\Furthermore, for the anticommuting  sector, we have the following propagators:
\begin{eqnarray}
\langle \omega^{ab}_{\mu}(p)\bar{\omega}^{cd}_{\nu}(-p)\rangle &=& -\frac{1}{p^{2}}\,\delta^{ac}\delta^{bd}\delta_{\mu\nu}\,,\\\cr
\langle c^{a}(p)\bar{c}^{b}(-p)\rangle &=&\frac{1}{p^{2}}\,\delta^{ab}\,,\\\cr
\langle \bar{c}^{a}(p)u^{b}(-p)\rangle &=&\frac{1}{p^{2}}\,\delta^{ab}\,,\\\cr
\langle \tau^{a}_{\mu}(p)\eta^{b}_{\nu}(-p)\rangle &=&-P_{\mu\nu}\,\delta^{ab}\,,\\\cr
\langle \tau^{a}_{\mu}(p)u^{b}(-p)\rangle &=&-i\frac{p_{\mu}}{p^{2}}\,\delta^{ab}\,,\\\cr
\langle \eta^{a}_{\mu}(p)\vartheta^{b}(-p)\rangle &=&-i\frac{p_{\mu}}{p^{2}}\,\delta^{ab}\,,\\\cr
\langle \vartheta^{a}(p)u^{b}(-p)\rangle &=&-\frac{1}{p^{2}}\,\delta^{ab}\,,\\\cr
\end{eqnarray}

All remaining propagators which have not been listed are vanishing.
\newpage 
\section{Tables of Quantum numbers}
We display here the  quantum numbers of all fields, sources and parameters of the model. In the following we shall employ the notation ``B'' for denoting the bosonic nature of a variable and ``F'' in the anticommuting case

{\small
$$
\begin{tabular}{|l|c|c|c|c|c|c|c|c|c|c|c|c|c|}
\hline
Fields & $A_{\mu}^{a}$ & $b^{a}$ & $\bar{c}^{a}$ & $c^{a}$ & $\bar{\varphi}^{a}_{i}$ & $\varphi^{a}_{i}$  & $B_{\mu}^{a}$ & $\bar{\omega}^{a}_{i}$ & ${\omega}^{a}_{i}$ & $\eta^{a}_{\mu}$ & $\xi^{a}$ & $u^{a}$ & $\phantom{\Big|}\!\lambda_{\mu}^{a}\!\phantom{\Big|}$  \\ \hline
Mass dimension & $1$ & $2$ & $2$ & $0$ & $1$ & $1$ & $1$ & $1$ & $1$ & $1$ & $0$ & $0$ & $3$ \\ \hline
Ghost number & $0$ & $0$ & $-1$ & $1$ & $0$ & $0$ & $0$ & $-1$ & $1$ & $1$ & $0$ & $1$  & $0$   \\ \hline
$q_{4(N^{2}-1)}$-Charge & $0$ & $0$ & $0$ & $0$ & $-1$ & $1$ & $0$ & $-1$ & $1$ & $0$ & $0$ & $0$ & $0$ \\ \hline
$\mathcal{Q}$-Charge & $0$ & $0$ & $0$ & $0$ & $0$ & $0$ & $0$ & $0$ & $0$ & $0$ & $0$ & $0$ & $1$  \\ \hline
Nature & B & B & F & F & B & B & B & F & F & F & B & F& B \\ \hline
\end{tabular}
$$
}

{\small
$$
\begin{tabular}{|l|c|c|c|c|c|c|c|c|c|c|c|}
\hline
Fields & $X^{ab}$ & $Y^{ab}$ & $\widetilde{X}^{ab}$ & $\widetilde{Y}^{ab}$ & ${T}^{ab}$ & $H^{ab}$ & $\widetilde{T}^{ab}$ & $\widetilde{H}^{ab}$ & $\rho^{a}$ & $\vartheta^{a}$  & $\phantom{\Big|}\!\tau_{\mu}^{a}\!\phantom{\Big|}$ \\ \hline
Mass dimension & $0$ & $0$ & $0$ & $0$ & $3$ & $3$ & $3$ & $3$ & $2$ & $2$   & $3$  \\ \hline
Ghost number & $0$ & $1$ & $0$ & $1$ & $-1$ & $0$ & $-1$ & $0$ & $0$ & $-1$  & $-1$   \\ \hline
$q_{4(N^{2}-1)}$-Charge & $0$ & $0$ & $0$ & $0$ & $0$ & $0$ & $0$ & $0$ & $0$ & $0$  & $0$  \\ \hline
$\mathcal{Q}$-Charge & $-1$ & $-1$ & $-1$ & $-1$ & $1$ & $1$ & $1$ & $1$ & $0$ & $0$  & $1$   \\ \hline
Nature & B & F & B & F & F & B & F & B & B & F & F   \\ \hline
\end{tabular}
$$
}

{\small
\centering
$$
\begin{tabular}{|l|c|c|c|c|c|c|c|c|c|}
\hline
Sources \& Parameters & $\bar{M}^{a}_{\mu i}$ & $M^{a}_{\mu i}$ & $N^{a}_{\mu i}$ & $\bar{N}^{a}_{\mu i}$ 	 &$\Omega_{\mu}^{a}$ & $L^{a}$ & $g$ &$\alpha$ & $\phantom{\Big|}\!\kappa\!\phantom{\Big|}$  \\ \hline
Mass dimension & $2$ & $2$ & $2$ & $2$ & $3$ & $4$ & $0$&0 & $0$  \\ \hline
Ghost number & $0$ & $0$ & $1$ & $-1$ & $-1$ & $-2$ & $0$&0 & $0$  \\ \hline
$q_{4(N^{2}-1)}$-Charge & $-1$ & $1$ & $1$ & $-1$ & $0$ & $0$ & $0$&0 & $0$ \\ \hline
$\mathcal{Q}$-Charge & $0$ & $0$ & $0$ & $0$ & $0$ & $0$ & $0$ &0 & $0$   \\ \hline
Nature & B & B & F & F & F & B & B & B & B \\ \hline
\end{tabular}
$$
}

{\small
\centering
$$
\begin{tabular}{|l|c|c|c|c|c|c|}
\hline
 Sources \& Parameters & $V^{abc}$ & $U^{abc}$ & $\Lambda$ & $\zeta_{1}^{abcdef}$ & $\zeta_{2}$ & $\phantom{\Big|}\!\zeta_{3}\!\phantom{\Big|}$  \\ \hline
Mass dimension & $2$ & $2$ & $2$ & $0$ & $0$ & $0$ \\ \hline
Ghost number & $0$ & $-1$ & $0$ & $0$ & $0$ & $0$   \\ \hline
$q_{4(N^{2}-1)}$-Charge & $0$ & $0$ & $0$ & $0$ & $0$ & $0$ \\ \hline
$\mathcal{Q}$-Charge & $0$ & $0$ & $0$ & $0$ & $0$ & $0$ \\ \hline
Nature & B & F & B & B & B & B \\ \hline
\end{tabular}
$$
}


\begin{thebibliography}{99}


\bibitem{Capri:2015ixa}
  M.~A.~L.~Capri {\it et al.},
  ``Exact nilpotent nonperturbative BRST symmetry for the Gribov-Zwanziger action in the linear covariant gauge,''
  Phys.\ Rev.\ D {\bf 92}, no. 4, 045039 (2015)
  doi:10.1103/PhysRevD.92.045039
  [arXiv:1506.06995 [hep-th]].
  
\bibitem{Capri:2016aqq} 
  M.~A.~L.~Capri {\it et al.},
  ``A local and BRST-invariant Yang-Mills theory within the Gribov horizon,''
  arXiv:1605.02610 [hep-th].

\bibitem{Gribov:1977wm} 
  V.~N.~Gribov,
  Nucl.\ Phys.\ B {\bf 139}, 1 (1978).
  doi:10.1016/0550-3213(78)90175-X

\bibitem{Sobreiro:2005ec} 
  R.~F.~Sobreiro and S.~P.~Sorella,
  hep-th/0504095.
	
\bibitem{Vandersickel:2012tz} 
  N.~Vandersickel and D.~Zwanziger,
  Phys.\ Rept.\  {\bf 520}, 175 (2012)
  doi:10.1016/j.physrep.2012.07.003
  [arXiv:1202.1491 [hep-th]].
		
\bibitem{Vandersickel:2011zc} 
  N.~Vandersickel,
  arXiv:1104.1315 [hep-th].
			
\bibitem{Pereira:2016inn} 
  A.~D.~Pereira,
  arXiv:1607.00365 [hep-th].
			
\bibitem{Singer:1978dk} 
  I.~M.~Singer,
  Commun.\ Math.\ Phys.\  {\bf 60}, 7 (1978).
  doi:10.1007/BF01609471
		
\bibitem{Dell'Antonio:1991xt} 
  G.~Dell'Antonio and D.~Zwanziger,
  Commun.\ Math.\ Phys.\  {\bf 138}, 291 (1991).
  doi:10.1007/BF02099494

\bibitem{Zwanziger:1989mf} 
  D.~Zwanziger,
  Nucl.\ Phys.\ B {\bf 323}, 513 (1989).
  doi:10.1016/0550-3213(89)90122-3
	
\bibitem{Capri:2012wx} 
  M.~A.~L.~Capri, D.~Dudal, M.~S.~Guimaraes, L.~F.~Palhares and S.~P.~Sorella,
  Phys.\ Lett.\ B {\bf 719}, 448 (2013)
  doi:10.1016/j.physletb.2013.01.039
  [arXiv:1212.2419 [hep-th]].

\bibitem{Sobreiro:2005vn}
  R.~F.~Sobreiro and S.~P.~Sorella,
  ``A Study of the Gribov copies in linear covariant gauges in Euclidean Yang-Mills theories,''
  JHEP {\bf 0506}, 054 (2005)
  doi:10.1088/1126-6708/2005/06/054
  [hep-th/0506165].

\bibitem{Capri:2015pja}
  M.~A.~L.~Capri, A.~D.~Pereira, R.~F.~Sobreiro and S.~P.~Sorella,
  ``Non-perturbative treatment of the linear covariant gauges by taking into account the Gribov copies,''
  Eur.\ Phys.\ J.\ C {\bf 75}, no. 10, 479 (2015)
  doi:10.1140/epjc/s10052-015-3707-z
  [arXiv:1505.05467 [hep-th]].



\bibitem{Capri:2015nzw}
  M.~A.~L.~Capri {\it et al.},
  ``More on the nonperturbative Gribov-Zwanziger quantization of linear covariant gauges,''
  Phys.\ Rev.\ D {\bf 93}, no. 6, 065019 (2016)
  doi:10.1103/PhysRevD.93.065019
  [arXiv:1512.05833 [hep-th]].


\bibitem{Cucchieri:2009kk} 
  A.~Cucchieri, T.~Mendes and E.~M.~S.~Santos,
  Phys.\ Rev.\ Lett.\  {\bf 103}, 141602 (2009)
  doi:10.1103/PhysRevLett.103.141602
  [arXiv:0907.4138 [hep-lat]].
		
\bibitem{Bicudo:2015rma} 
  P.~Bicudo, D.~Binosi, N.~Cardoso, O.~Oliveira and P.~J.~Silva,
  Phys.\ Rev.\ D {\bf 92}, no. 11, 114514 (2015)
  doi:10.1103/PhysRevD.92.114514
  [arXiv:1505.05897 [hep-lat]].

\bibitem{Dudal:2008sp} 
  D.~Dudal, J.~A.~Gracey, S.~P.~Sorella, N.~Vandersickel and H.~Verschelde,
  Phys.\ Rev.\ D {\bf 78}, 065047 (2008)
  doi:10.1103/PhysRevD.78.065047
  [arXiv:0806.4348 [hep-th]].
	
\bibitem{Capri:2015pfa} 
  M.~A.~L.~Capri, D.~Fiorentini and S.~P.~Sorella,
  Phys.\ Lett.\ B {\bf 751}, 262 (2015)
  doi:10.1016/j.physletb.2015.10.032
  [arXiv:1507.05481 [hep-th]].
	
\bibitem{Guimaraes:2015bra} 
  M.~S.~Guimaraes, B.~W.~Mintz and S.~P.~Sorella,
  Phys.\ Rev.\ D {\bf 91}, no. 12, 121701 (2015)
  doi:10.1103/PhysRevD.91.121701
  [arXiv:1503.03120 [hep-th]].

\bibitem{Aguilar:2015nqa} 
  A.~C.~Aguilar, D.~Binosi and J.~Papavassiliou,
  Phys.\ Rev.\ D {\bf 91}, no. 8, 085014 (2015)
  doi:10.1103/PhysRevD.91.085014
  [arXiv:1501.07150 [hep-ph]].

\bibitem{Huber:2015ria} 
  M.~Q.~Huber,
  Phys.\ Rev.\ D {\bf 91}, no. 8, 085018 (2015)
  doi:10.1103/PhysRevD.91.085018
  [arXiv:1502.04057 [hep-ph]].
	
		
			
\bibitem{Siringo:2014lva} 
  F.~Siringo,
  Phys.\ Rev.\ D {\bf 90}, no. 9, 094021 (2014)
  doi:10.1103/PhysRevD.90.094021
  [arXiv:1408.5313 [hep-ph]].

\bibitem{Machado:2016cij} 
  F.~A.~Machado,
  arXiv:1601.02067 [hep-ph].

\bibitem{Moshin:2015gsa} 
  P.~Y.~Moshin and A.~A.~Reshetnyak,
  Physics A {\bf 31}, 1650111 (2016)
  doi:10.1142/S0217751X16501116
  [arXiv:1506.04660 [hep-th]].

\bibitem{Piguet:1995er} 
  O.~Piguet and S.~P.~Sorella,
  Lect.\ Notes Phys.\ M {\bf 28}, 1 (1995).


\bibitem{Fiorentini:2016rwx} 
  M.~A.~L.~Capri, D.~Fiorentini, M.~S.~Guimaraes, B.~W.~Mintz, L.~F.~Palhares and S.~P.~Sorella,
  ``A local and renormalizable framework for the gauge-invariant operator $A^2_{\min}$ in Euclidean Yang-Mills theories in linear covariant gauges,''
  arXiv:1606.06601 [hep-th].
  
  \bibitem{progress}
  M.~A.~L.~Capri {\it et al.}, work in progress. 

\end{thebibliography}
\end{document}